\documentclass[lettersize,journal]{IEEEtran}
\usepackage{amsmath,amsfonts}
\usepackage{algorithmic}
\usepackage{array}
\usepackage[caption=false,font=normalsize,labelfont=sf,textfont=sf]{subfig}
\usepackage{textcomp}
\usepackage{stfloats}
\usepackage{url}
\usepackage{verbatim}
\usepackage{graphicx}
\usepackage{threeparttable}
\usepackage{xcolor}
\usepackage[symbol]{footmisc}
\usepackage{hhline}
\usepackage{multirow}
\def\BibTeX{{\rm B\kern-.05em{\sc i\kern-.025em b}\kern-.08em
    T\kern-.1667em\lower.7ex\hbox{E}\kern-.125emX}}
\usepackage{balance}
\begin{document}
\title{A 399$\mu$W 114.3 dB DR Companding Readout ASIC for MEMS Microphones Employing a Multirate Time-Domain ADC}
\author{Javier Granizo\IEEEauthorrefmark{1},~\IEEEmembership{Student Member,~IEEE}, Ruben Garvi\IEEEauthorrefmark{1}, Ricardo Carrero\IEEEauthorrefmark{1}, Jorge de la Torre\IEEEauthorrefmark{1}, Javier Fernandez\IEEEauthorrefmark{1}, Dietmar Straeussnigg\IEEEauthorrefmark{2}, Andreas Wiesbauer\IEEEauthorrefmark{2},~\IEEEmembership{Member,~IEEE}, and Luis Hernandez\IEEEauthorrefmark{1},~\IEEEmembership{Senior Member,~IEEE}%
\thanks{This research was funded by grant FPU21/02257 of the Spanish Ministry of Science, Innovation and Universities.\\
\IEEEauthorrefmark{1} Carlos III University, Leganes, Spain.  
\IEEEauthorrefmark{2} Infineon Technologies Austria Ag.}}

\markboth{Journal of \LaTeX\ Class Files,~Vol.~18, No.~9, June~2026}%
{How to Use the IEEEtran \LaTeX \ Templates}

\maketitle

\begin{abstract}
Improvements in the dynamic range and sensitivity of digital MEMS microphones are essential in  applications like advanced noise canceling and voice recognition. A cost effective solution to achieve these goals is the companding ADC architecture. Companding ADCs split the dynamic range in several segments with different quantization noise levels, relaxing power constraints. A common problem of companding microphones are audible artifacts generated when the input signal crosses the boundaries between different amplitude segments. We show in this paper a companding ADC architecture that mitigates the boundary artifacts by leveraging the instantaneous and high-resolution time-domain representation of the input signal in a VCO-based ADC. The use of a multi-rate frequency-to-digital converter allows to decouple quantization noise from the VCO frequency, keeping standard audio sampling rates. Co-optimization of the driver and oscillator circuits enables our VCO-ADC to reach \textgreater 112dBc of peak SFDR without a feedback DAC, keeping a Giga-Ohm input impedance compatible with a capacitive MEMS. We show measurements of a 0.13 $\mu$m ASIC implementing a complete readout circuit for a digital MEMS microphone. This includes two analog channels and the digital signal processing and calibration blocks required to deliver a standard single-bit PDM output. This ADC reaches a dynamic range of 114.3dB with a power budget under 400 $\mu$W, a Schreier FoM$_{SNDR}$ of 171.0 dB and a FoM$_{DR}$ of 191.3 dB.

\end{abstract}


\section{Introduction}
\begin{figure}
    \centering
    \includegraphics[width=\linewidth]{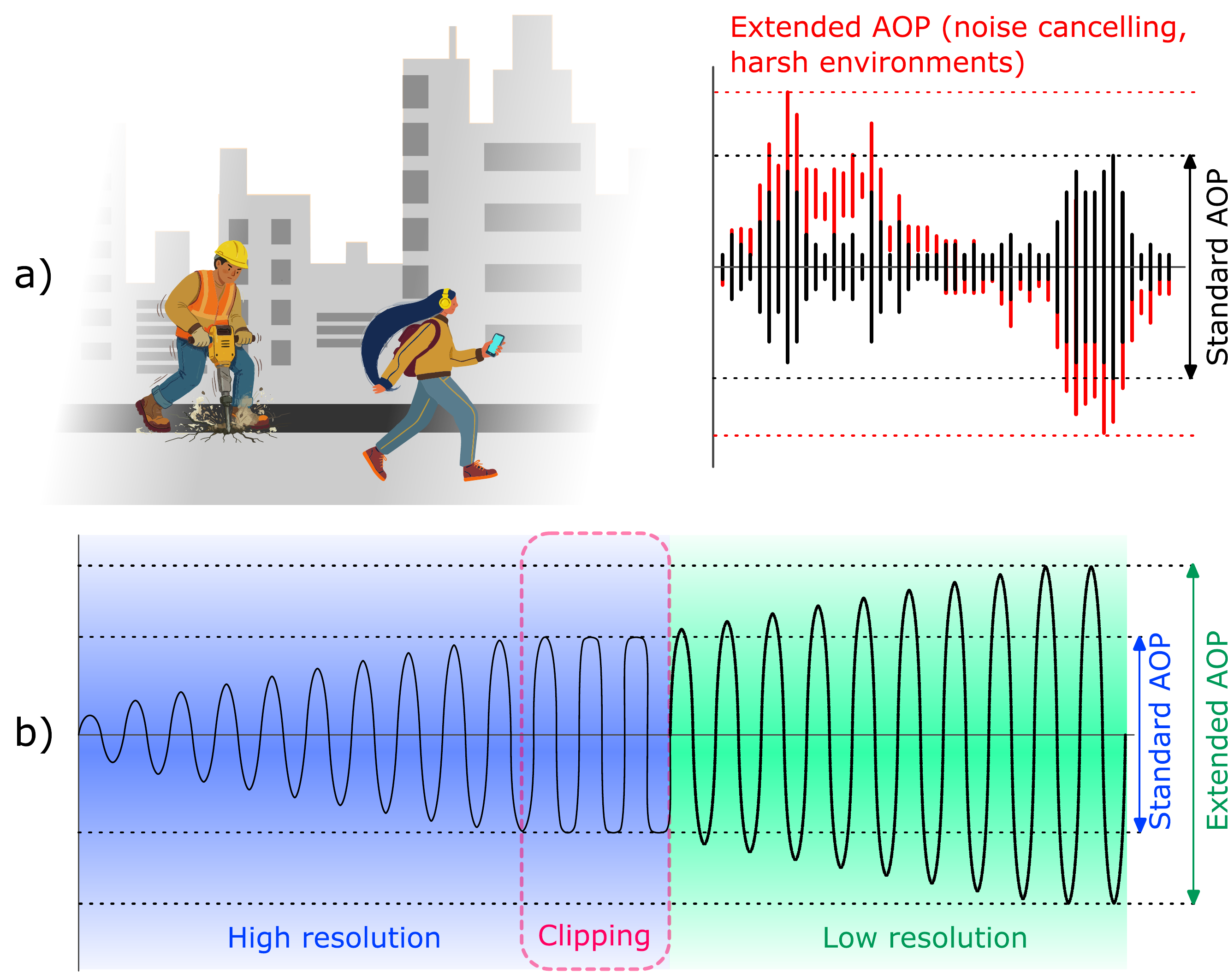}
    \caption{(a) Example of application of companding MEMS microphones.(b) Dynamic range segmentation in companding ADCs.}
    \label{fig:companding_mismatch_problem}
\end{figure}

\IEEEPARstart{D}{ynamic} range is a key parameter of the digital MEMS microphones used in consumer and automotive applications. This Dynamic Range (DR) is defined in audio as the signal span between the threshold level of 0dB of SNR and the Acoustic Overload Point (AOP). The AOP is commonly defined as the Sound Pressure Level (SPL) reaching a 10\% of Total Harmonic Distortion (THD). AOP values in excess of 130dBSPL are required to guarantee a proper handling of audio signals in harsh acoustic environments such as public transport, construction sites and airplanes. Fig. \ref{fig:companding_mismatch_problem}(a) shows an example of one of such scenarios, where we want to digitize a voice signal in an environment with a loud noise that has to be canceled by a noise-cancelling algorithm.

A hardware-efficient implementation of such extended-range microphones uses a companding ADC \cite{basic_companding}, where the DR is split in several segments. Each segment encodes the signal with a different quantization step which results in a signal-dependent quantization noise. High volume sounds are encoded with high quantization noise, but human hearing is less sensitive to noise for loud sounds, which ideally makes companding barely noticeable. The implementation benefit is that an ADC with a moderate dynamic range can be employed if it is preceded by a Programmable Gain Amplifier (PGA) whose gain is adapted according to the signal and succeeded by a digital gain correction stage \cite{Ceballos_companding}. However, when the input signal is close to the transitions between different segments, an undesirable effect happens (see Fig. \ref{fig:companding_mismatch_problem}(b)). This effect consists of audible glitches. One cause of the glitches is the gain and offset mismatch in the transitions between DR segments. Another source of audible glitches is the latency between the detection of the input amplitude and the change in the DR segment,  which influences the duration of the clipping time in Fig. \ref{fig:companding_mismatch_problem}(b). Therefore two key hardware requirements of the companding system are an excellent matching between the PGA in front of the ADC and the digital gain correction stage, and a fast and accurate estimation of the signal amplitude. Even under ideal conditions, a third mechanism may produce audible glitches depending on the ADC architecture, especially when it has memory as in a sigma-delta modulator. 

Many digital MEMS microphones use Sigma-Delta Modulators (SDM) as ADCs \cite{Malcovati_MEMS_microphone_DSM, pavan_MASH_CTSDM, makinwa, youngcheol}. These modulators produce a noise shaped signal with moderate resolution ($\leq$ 4 bits) but a high order noise shaping to reach a large dynamic range. Solutions incorporating companding techniques have been proposed to extend the dynamic range of switched capacitor sigma-delta modulators. As an example, in \cite{Ceballos_companding, companding_100ms} a programmable preamplifier is used in front of a single switched capacitor sigma-delta modulator, whose output is scaled digitally afterwards. In other MEMS microphones, the outputs of several sigma delta modulators operating in parallel are combined to produce a higher dynamic range response \cite{perrott2025adaptive}. All these solutions require a sort of decimation filter included in the microphone that extracts an accurate time-domain representation of the input signal. This representation is needed to change the gain of the preamplifier or select the output of the modulator according to the input signal level. As a consequence, DR segment change cannot be fast due to the delay introduced by the decimation filter. The standard output of MEMS microphones uses a Pulse Density Modulation (PDM) format, making the decimation filter unnecessary except for handling the companding mechanism. On top of that, solutions using a programmable gain preamplifier suffer from the slow response time of the preamplifier and different offset levels when gain is changed. All these implementation problems challenge the design of companding MEMS microphones with conventional SDMs. 

A different implementation approach for digital MEMS microphones uses time-domain ADC techniques. The ADCs described in \cite{Quintero_VCOADC, Carlos_DOC1, Carlos_mls, Medina_2nd}, have shown that open-loop Voltage Controlled Oscillator ADCs (VCO-ADCs) \cite{magazine_colorines_pt1}, despite of the supposed lack of linearity of Ring Oscillators (RO), can be used to implement digital MEMS microphones with improved area and power compared to switched capacitor designs. The efficiency of open loop VCO-ADCs relies in the possibility to reduce the driving circuit between the MEMS and the VCO to a single-transistor source follower or transconductor \cite{why_and_how}. In addition, VCO-ADCs provide a first-order noise shaped output but with word-lengths in excess of 9 bits, providing an immediate and accurate time-domain representation of the input signal without any extra filtering. We show in this paper how these two features make VCO-ADCs a good option for MEMS microphones using companding. 

To the author's knowledge, the first companding VCO-ADC has been described in \cite{DOGX_ESSERC}, which is the start point of the chip described in the present paper. In \cite{DOGX_ESSERC}, we show two new concepts, firstly the companding architecture using two VCO-ADCs and secondly a VCO-ADC topology that uses multiple sampling rates and a decimation filter to optimize the VCO noise and power. In the present paper, we show an improved version of our previous chip including an internal non-uniform clock generator and all the digital signal processing blocks to implement the ADC calibration and PDM output. We also show a system-level analysis of different companding architectures based on sigma-delta modulators, comparing the switching noise folding phenomena. Additionally, we discuss a theoretical explanation of why a periodically non-uniform sampling scheme does not compromise the SQNR, based on the PFM model of VCO-based quantizers \cite{gutierrezPulseFrequencyModulation2018}. The paper is structured as follows: Section II compares VCO-ADCs with other solutions to build an audio companding ADC. Section III introduces the periodic non-uniform sampling VCO-ADC architecture. Section IV explains the design of the chip. Section V shows the experimental measurements and Section VI concludes the paper. 

\section{Analysis of digital MEMS microphone ADCs with companding}

\begin{figure}[t]
\centering
{\includegraphics [width=\columnwidth]{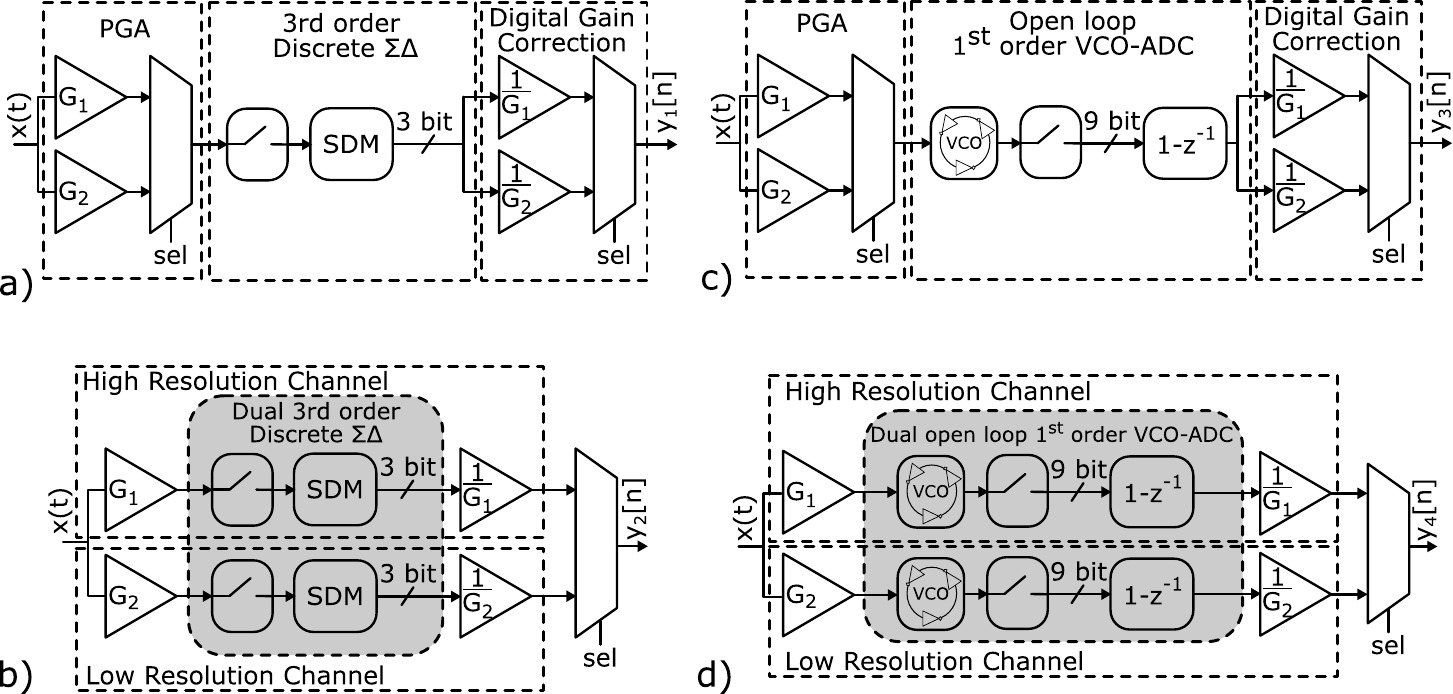}}
\caption{System-level comparison between different companding architectures for MEMS applications using SDMs and Open-loop VCO-ADCs:  (a) Companding SDM with a PGA. (b) Companding SDM employing two complete channels. (c) Companding open-loop VCO-ADC employing a PGA. (d) Companding open-loop VCO-ADC using two complete channels.} 
\label{fig:system_comparison_schematics}
\end{figure} 

\begin{figure}[t]
\centering
{\includegraphics [width=\columnwidth]{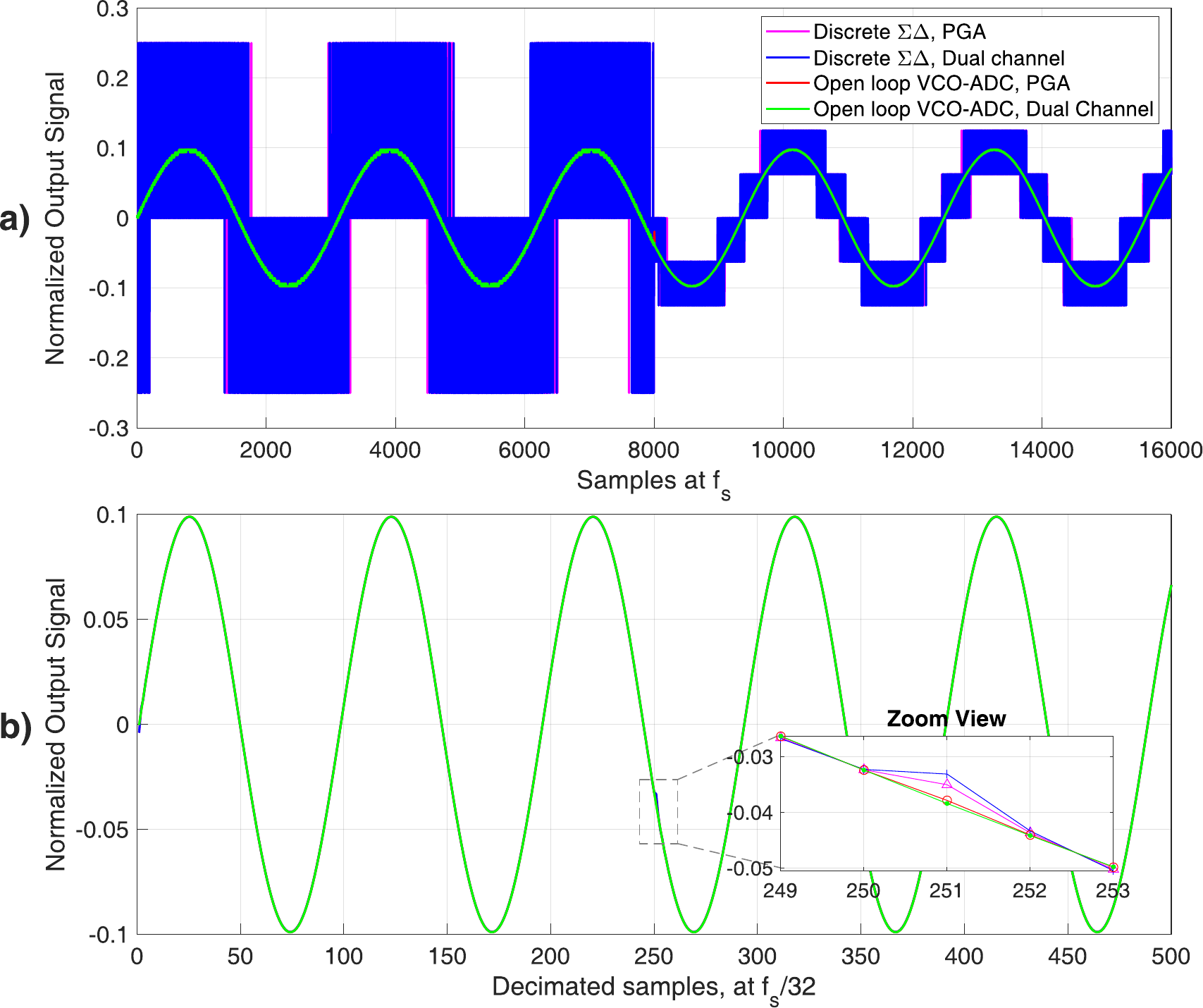}}
\caption{System-level simulations at -20dBFS comparing different companding architectures: (a) Normalized output signal showing the difference in quantization levels. (b) Normalized output signal after decimation.} 
\label{fig:system_comparison_transient}
\end{figure} 

\begin{figure}[t]
\centering
{\includegraphics [width=\columnwidth]{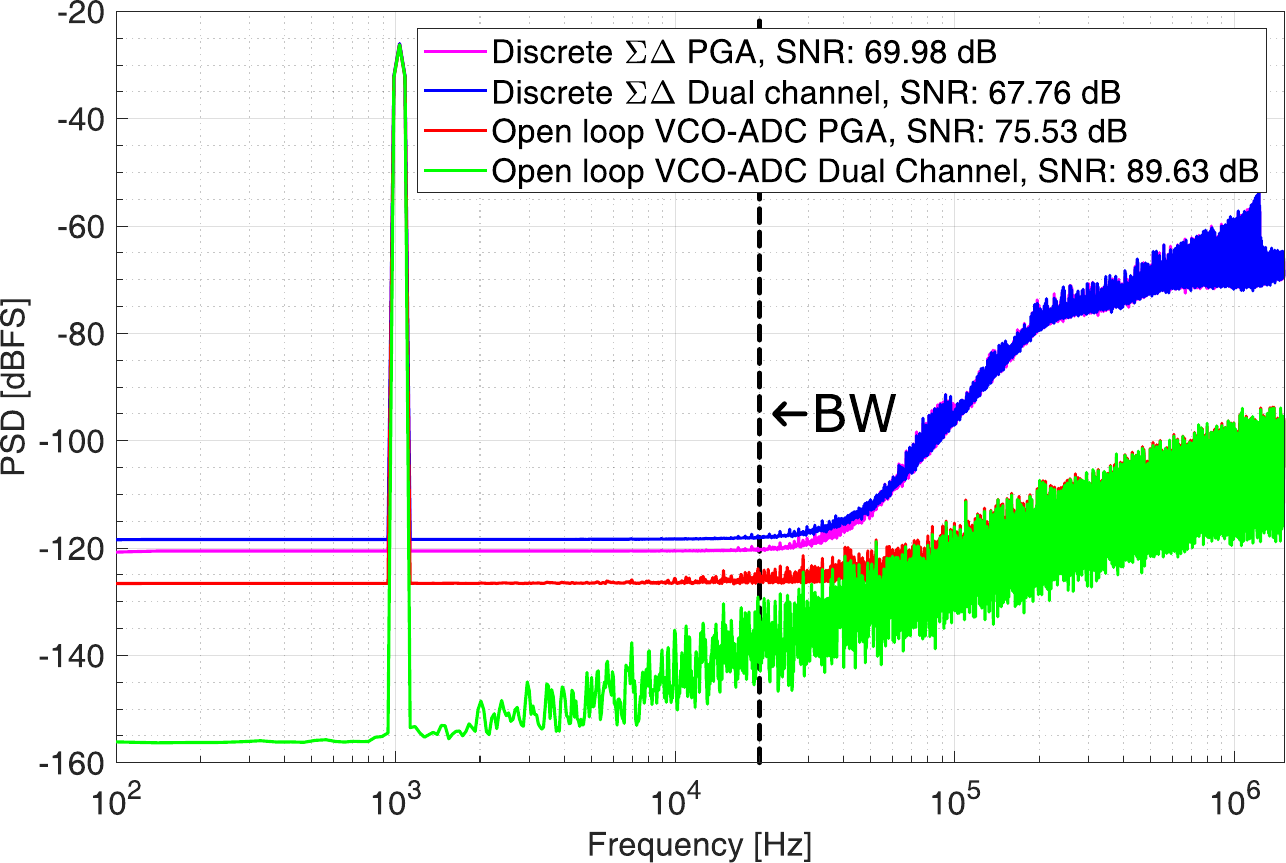}}
\caption{Averaged PSD of system-level simulations at -20dBFS comparing different companding architectures.} 
\label{fig:system_comparison_PSD}
\end{figure}

In this section, a brief system-level comparison between different ADC companding structures usable for a digital MEMS microphone is portrayed. We will focus in the potential perturbations introduced by the intrinsic ADC topology in the DR switching points instead of gain, offset or timing related issues. The different ADC topologies to be studied are shown in Fig. \ref{fig:system_comparison_schematics}. In Fig. \ref{fig:system_comparison_schematics}(a), a SDM companding implementation is shown using a PGA.  In  Fig. \ref{fig:system_comparison_schematics}(b), a companding two SDM channel topology is shown. In both cases, the SDM converter structure and coefficients are extracted from \cite{DSM_YumingLI_3order_CIFF_modulator}, as they fit the expected parameters for a MEMS microphone ADC in terms of power, DR and sampling frequency. The SDM of \cite{DSM_YumingLI_3order_CIFF_modulator} uses a cascade-of-integrators feedforward (CIFF) 3$^{rd}$ order analog loop filter and a 3-bit quantizer. Similar structures using 3$^{rd}$ order noise shaping with 3-4 bit quantizers are commonly found in the state-of-the-art of MEMS microphones \cite{Sensors_Luca_Sant_SSF_microphone,Malcovati_MEMS_microphone_DSM}. In Fig. \ref{fig:system_comparison_schematics}(c) and Fig. \ref{fig:system_comparison_schematics}(d), two companding VCO-ADC architectures are shown: one employing a PGA, and another using a two channel configuration. These VCO-ADC topologies employ a 1$^{st}$ order open-loop architecture, similar to the one shown in \cite{Carlos_DOC1}.
The main advantage of VCO-ADCs in comparison with conventional SDMs is that VCO quantizers can be easily implemented with \textgreater 9 bits. This is because in, a VCO-ADC, the quantizer resolution is dependent not only on the number of taps of the ring oscillator, but also on the ratio between the oscillator frequency and the sampling frequency \cite{magazine_colorines_pt1}. To establish a fair comparison between the architectures of Fig. \ref{fig:system_comparison_schematics}, the quantizer resolution of the VCO-ADC architectures has been scaled to 9 bits, resulting in a  simulated SQNR of 108dB at full scale which exactly matches with the DR of the switched capacitor SDM counterparts \cite{DSM_YumingLI_3order_CIFF_modulator}.

In Fig. \ref{fig:system_comparison_transient}, a transient simulation showcasing the companding systems of Fig. \ref{fig:system_comparison_schematics} is displayed. From this point and on, we will refer to the change from one analog gain (G$_1$) to another (G$_2$) by means of the $sel$ programming bit as a change in the mode (or channel) of the converter. This is independent of the companding architecture used, being either a dual-channel topology or using a PGA. In this simulation, all systems suffer a single transition from the low-resolution mode to the high-resolution mode. In the case of the PGA, the simulations consider an ideal PGA with infinite bandwidth, resulting in an instantaneous switch between modes. The in-band signal gain of the four case studies is fixed to 1, independently of the mode used. In other words, when switching from the low-resolution mode to the high-resolution mode, the quantization noise floor decreases proportionally to the analog gain difference (G$_1$/G$_2$). The simulations are performed with a 1kHz input tone at -20dBFS, and an analog signal gain difference of 4, resulting in a 12dB difference in the quantization noise floor and ensuring no ADC mode is saturated. The sampling frequency (f$_s$) of the systems is 3 MHz, the effective oscillation frequency of the VCO-ADC converters (f$_{ef}$) is 768 MHz \cite{magazine_colorines_pt1}, and their gain (K$_{vco}$) is 768 MHz/V. The VCO-ADC converters are configured in a pseudo-differential configuration, as commonly used in the literature \cite{Carlos_DOC1,Saux_MASH_rombouts}. Fig. \ref{fig:system_comparison_transient}(a) shows the difference between quantization steps when switching modes/channels, both in the SDM converters and in the open-loop VCO-ADCs. The smaller LSB value in the VCO-ADC provides a fine and clear representation of the input tone, despite having the same SQNR as the SDM converter. From this, a VCO-ADC can be advantageously used as a companding converter, as it is possible to quickly infer the input amplitude using a digital comparator without a decimation filter. This simplifies the decision logic to change between modes. It also decreases the latency to decide whether the input signal has surpassed a given threshold. This latency decrease proves useful in two ways: first, it removes clipping artifacts at frequencies close to the audio bandwidth (20kHz); and secondly, it can quickly detect if the input signal is high enough to produce significant distortion in the high-resolution mode. 

However, Fig. \ref{fig:system_comparison_transient}(a) does not clearly show if an error is produced in the output signal when switching between modes (it can be seen faintly in the open loop VCO-ADC with PGA, but not in any other case). To clearly represent whether a transient error happens when switching modes, Fig. \ref{fig:system_comparison_transient}(b) provides the same output signals as \ref{fig:system_comparison_transient}(a) after being processed by an $8^{th}$ order decimation filter with a  decimation factor of 32. As seen in the Fig. \ref{fig:system_comparison_transient}(b) zoom view, the mode switch results in an appreciable transient signal error in the case of the SDM converters. Nevertheless, the magnitude of this transient error depends both in the input signal being digitized, as well as on the internal states of the oversampling converter when switching between modes. A single transient simulation switching at a single point in time does not correctly portray the expected energy of such an error. 

To properly compare which architectures of Fig. \ref{fig:system_comparison_schematics} result in a lower-on-average transient error when switching, a different test has been performed. This test consists of averaging 128 transient simulations of Fig. \ref{fig:system_comparison_transient} changing randomly the instant at which the modes are switched. The average PSD of the output signal of the different companding systems is calculated, and represented in Fig. \ref{fig:system_comparison_PSD} for each companding architecture. Observing this figure, we can draw some conclusions regarding what architectures seem more tolerant to switching errors. In both of the SDM companding architectures we see a significant degradation of the noise floor of the converter. The nature of this error resides in the out-of-band quantization noise of the SDM converter being folded back to the pass-band when performing the mode switch. In the case of using an open-loop VCO-ADC with a PGA, despite the significant decrease in the out-of-band noise folding, the in-band noise floor is still degraded (although less than in the SDM cases). The authors theorize that the ideal switch performed by the PGA still results in an error due to the internal states of the VCO-ADC not being changed consequently when the switching happens. Nevertheless, in the case of a dual-channel open-loop VCO-ADC topology, the switching in-band error is mostly removed. For this reason, this topology has been chosen to implement the companding microphone ASIC described in this paper.

\section{VCO-ADC architectures for MEMS microphones}

\begin{figure}[t]
\centering
{\includegraphics [width=0.85\columnwidth]{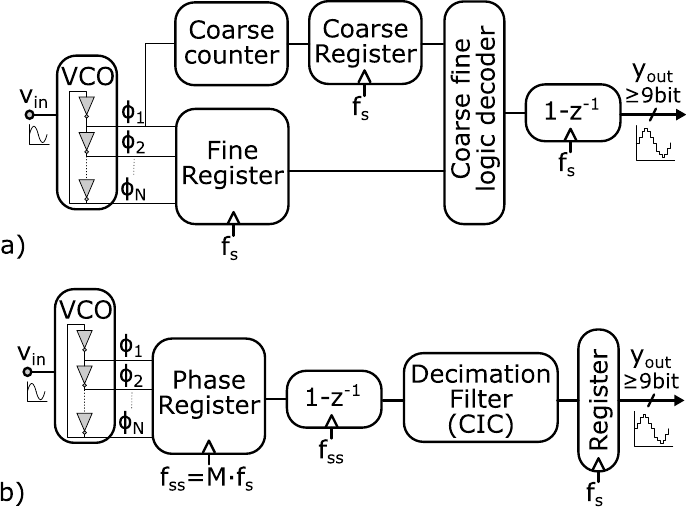}}
\caption{(a) Coarse-Fine VCO-ADC architecture. (b) Multi-rate sampling and decimator VCO-ADC architecture.} 
\label{fig:CF_vs_DLLCIF}
\end{figure}

Commercial MEMS microphones are bound to some interface restrictions to keep compatibility with legacy microphones using switched capacitor SDMs. These restrictions consist of a set of standard sampling rates typically between 1MHz and 5MHz and a single-bit PDM data interface. The DR required, considering quantization noise only, lies above 100dB. Implementing a $1^{st}$ order SDM with a VCO-ADC at such sampling frequencies requires counting and sampling the pulses produced by a RO with an effective frequency  ( 2 x oscillation frequency x number of phases, \cite{Carlos_DOC1, Optimization_Rombouts}) on the order of $f_{ef}=1GHz$. A power efficient method of implementing this asynchronous sampling and counting function is the coarse-fine architecture shown in Fig. \ref{fig:CF_vs_DLLCIF}(a) and described for instance in \cite{Carlos_DOC1, Huang_Mercier_VCO_audio,watanabe_TAD_16nm, dehaene_VCO_ADC, Vesterbaka_sampling_error_VCO_ADC, moulin_arbiter}. After the frequency-to-digital conversion, a digital noise shaper re-encodes the VCO-ADC multibit output into a single-bit, PDM signal. In spite of this power efficient frequency-to-digital conversion, the VCO effective frequency is dictated by quantization noise considerations. However, a power saving along with better performance can be achieved if the VCO frequency can be chosen to optimize the VCO phase noise. The phase noise of the VCO is what ultimately determines the Input Referred Noise (IRN) of the VCO-ADC \cite{cardesSNDRLimitsOscillatorBased2018}. If we are not bound to quantization noise restrictions, we can afford a lower VCO oscillation frequency, which allows larger RO inverter transistors, improving flicker noise without increasing the number of phases \cite{abidi_phase_noise}. Considering that the VCO-ADC requires either level shifters or sense amplifiers \cite{staszewsky_speed} to eliminate the amplitude modulation conveyed in the RO outputs, a reduction of the number of phases and VCO frequency may improve the power and area over coarse-fine architectures.

The VCO-ADC architecture displayed in Fig. \ref{fig:CF_vs_DLLCIF}(b) achieves the goal of decoupling the VCO effective frequency from the microphone sampling frequency $f_{s}=1/T$ by introducing a first sampling stage at frequency $f_{ss}=M \cdot f_{s}$. Integer factor M is chosen such that the required SQNR$_{peak}$ of the microphone is achieved with the VCO effective frequency $f_{ef}$ that best fits the desired VCO IRN, power and area. Assuming a signal bandwidth $BW$ and a differential configuration, \eqref{eq:dr_DLL_cif} \cite{Optimization_Rombouts} predicts a SQNR enhancement by a factor $10 \cdot log_{10}(M)$:

\begin{eqnarray}
SQNR_{peak}=-3.41dB + 20 log_{10}(f_{ef}) + \nonumber  \\
 + 10 log_{10}({f_s}) - 30 log_{10}({BW}) + 10 log_{10}(M) 
\label{eq:dr_DLL_cif}
\end{eqnarray}

To adapt the sampling rate $f_{ss}$ to an industry standard sampling rate $f_{s}$, we require a decimation filter \cite{DOGX_ESSERC}. This decimation filter can implemented with a Cascade of Integrators Comb (CIC) filter \cite{laddomada}. Considering that the VCO-ADC produces $1^{st}$ order noise shaping, the low-pass decimation function of the CIC filter must be at least of $2^{nd}$ order to avoid SQNR degradation. The architecture depicted in \ref{fig:CF_vs_DLLCIF}(b) was successfully used in the companding VCO-ADC described in \cite{DOGX_ESSERC}. However in that case, frequency $f_{s}$ was generated by a divider by M=8 from an external clock $f_{ss}=24MHz$. In a practical case, $f_{s}$ is provided externally to the microphone and generation of $f_{ss}$ has to be done in the microphone. The chip described in this paper is a complete microphone ASIC including this functionality, as a difference to \cite{DOGX_ESSERC}.

\begin{figure}[t]
\centering
{\includegraphics [width=\columnwidth]{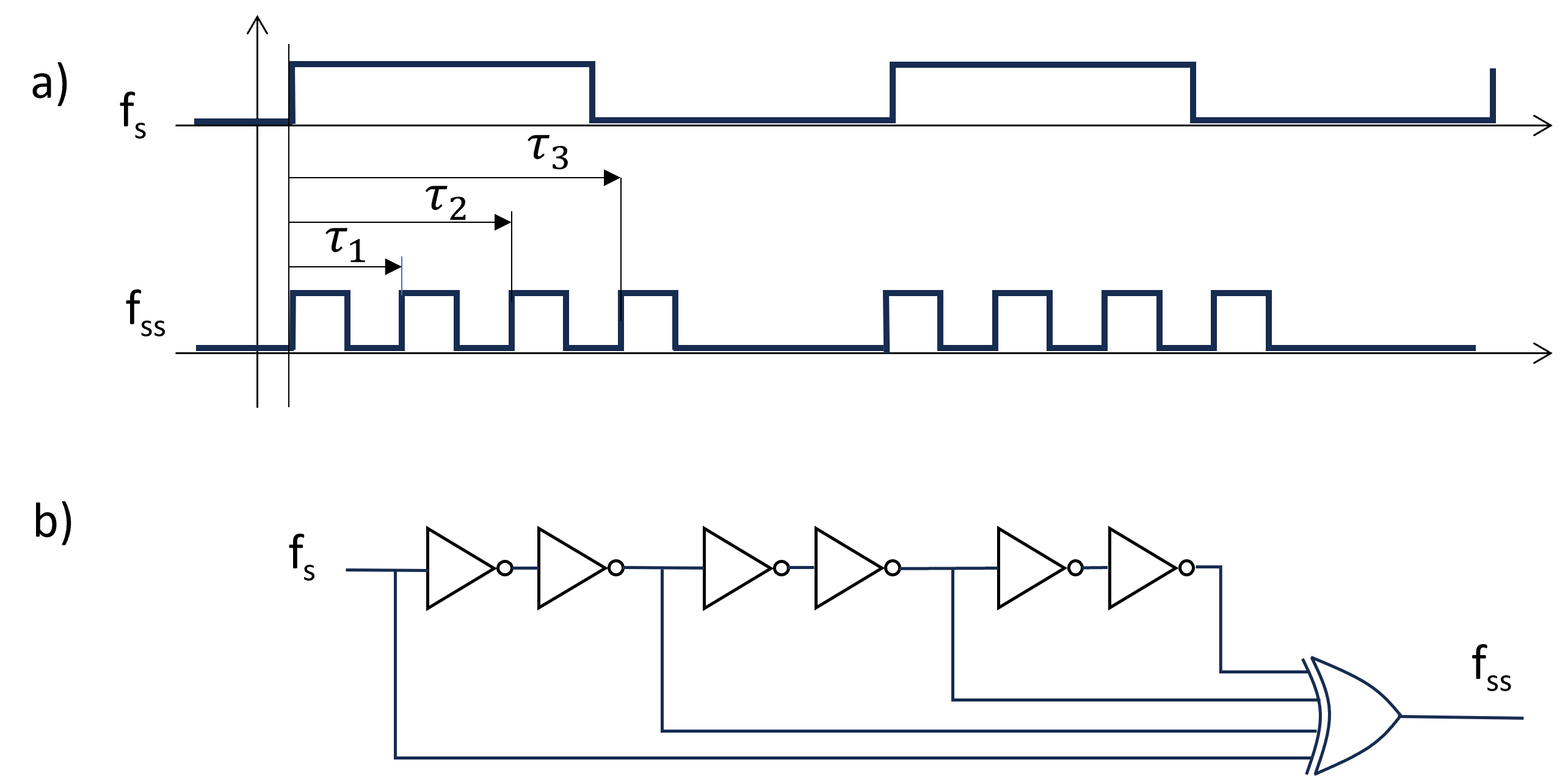}}
\caption{(a) Periodically non-uniform sampling clock (b) Implementation circuit example.} 
\label{fig:XOR_DLL}
\end{figure}

To generate a fast clock at frequency $f_{ss}$, the external clock $f_{s}$ can be multiplied, using for instance a PLL. However, analyzing the multi-rate VCO-ADC architecture, it has been found that we can afford a non-uniform distribution of the sampling pulses in $f_{ss}$, as long as its sequence is periodically repeated at frequency $f_{s}$. This situation is depicted in Fig. \ref{fig:XOR_DLL}(a), where a sequence of four pulses in $f_{ss}$, unevenly delayed by times $\tau_1...\tau_3$, is repeated every clock cycle in $f_{s}$. For instance, such non-uniform sampling clock can be generated by a chain of digital inverter delays (see Fig. \ref{fig:XOR_DLL}(b)), allowing to derive $f_{ss}$ from $f_{s}$, with an open loop, simpler  circuit than a PLL.

\begin{figure}[t]
\centering
{\includegraphics [width=0.95\columnwidth]{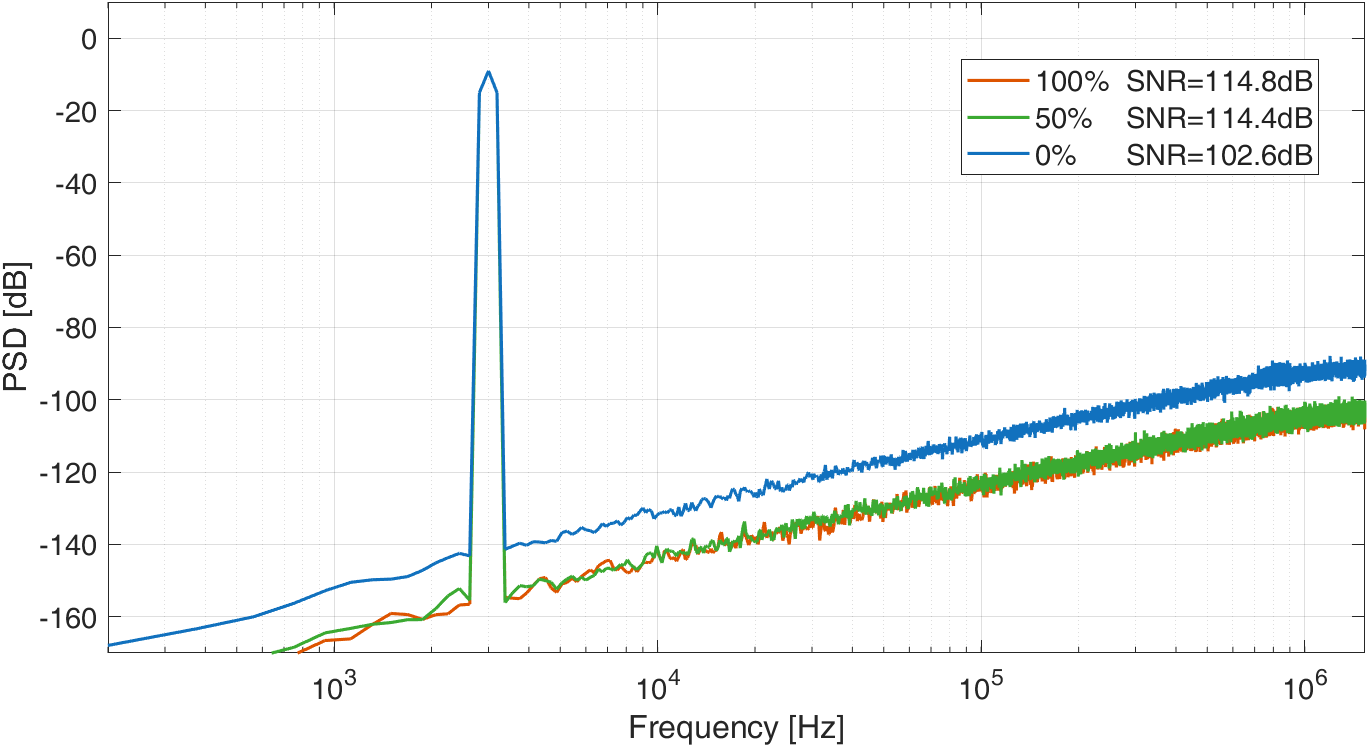}}
\caption{FFT of time domain simulation with variable delay time errors} 
\label{fig:sim_gap}
\end{figure}

As a verification of this simplification, Fig. \ref{fig:sim_gap} shows the FFT of a transient simulation of Fig. \ref{fig:CF_vs_DLLCIF}(b) using the clock generator of Fig. \ref{fig:XOR_DLL}(b) with $f_s=3MHz$, M=16, $f_{ef}=768MHz$. The delays in the delay chain will be assumed equal and with delays $\tau_k$, proportionally reduced compared with the nominal values $\tau_k=k \cdot T/M$. This is a hardware-wise plausible situation, as most inverters in Fig. \ref{fig:XOR_DLL}(b) will have little mismatch among them but large differences with their nominal values due to Process, Temperature and Voltage (PVT) variations. Green and red lines have $100\%$ and $50 \%$ of the nominal delays producing 114.8dB and 114.4dB of SQNR respectively for a $-6dB_{FS}$  input. The blue line corresponds to $f_{ss}=f_s$ and has 102.6dB of SQNR confirming \eqref{eq:dr_DLL_cif}. As can be observed, a large error in the sampling points $\tau_k$ produces a negligible SQNR difference up to the point where all delays collapse into a single sampling instant. This circuit simplification plays a substantial role in preferring the multi-rate architecture over the coarse-fine architecture and may be leveraged in other application fields as biomedical signals \cite{Pochet_Mercier_Hall_HiZ_VCO} or high-speed data converters. A detailed mathematical explanation of this phenomena can be found in Appendix A.

\section{ASIC implementation}
\begin{figure}
    \centering
    \includegraphics[width=\linewidth]{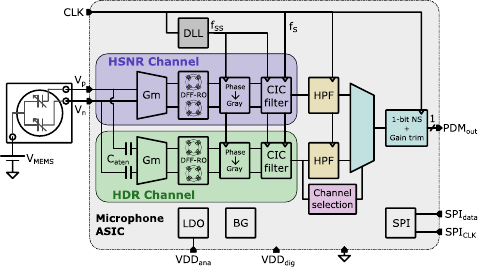}
    \caption{Proposed companding VCO-ADC system-level overview.}
    \label{fig:full_sys_level}
\end{figure}

Based on the analysis performed in Sections II and III of this paper, the companding VCO-ADC architecture described in Fig. \ref{fig:full_sys_level} has been implemented in a proof-of-concept ASIC using 0.13$\mu$m CMOS process. This companding ADC implements a dual-channel topology (Fig. \ref{fig:system_comparison_schematics}(d)), where the high resolution channel is named high-signal-to-noise-ratio (HSNR) channel, and the low resolution channel is named high-dynamic-range (HDR) channel. Each channel includes a transconductor (GM), a differential feed-forward ring oscillator (DFF-RO) \cite{borgmans_analog_VCO}, a phase-to-gray encoder \cite{ DOGX_ESSERC, medinaGrayEncodedRingOscillator2023}, a gray-to-binary decoder, and a CIC filter. The phase-to-gray encoder, gray-to-binary decoder and CIC filter are driven by a delay line (f$_{ss}$), acting as the clock multiplier described in Section III. This delay line is supervised by a digital DLL circuit with a dead zone. The CIC filter performs the decimation back to a standard sampling rate supplied to the ASIC. The HSNR channel presents x4 times higher analog gain (K$_{vco}$) than the HDR channel. The HDR channel compensates for this discrepancy in the gain by scaling its quantized digital signal by 4 after the CIC filter. The channel selection logic selects the use of the HSNR channel or the HDR channel depending on the input amplitude level. To do so, it uses the HDR channel output to obtain a sufficiently high-resolution, undistorted digital representation of the input signal. After the channel selection, the channel is further processed to a 1-bit, PDM-compatible output signal by means of a digital noise-shaper (NS). In order to ensure minimal audible glitches, the offset of the channels is high-pass filtered (HPF) and the 1-bit NS performs fine gain digital calibration (for inter-channel gain differences under $\leq$0.4dB). Additionally, the ASIC integrates a band-gap reference (BG), a supply voltage regulator (LDO), and a serial peripheral programming interface (SPI). These blocks are standard circuits \cite{Ceballos_companding,Malcovati_MEMS_microphone_DSM,companding_100ms,ceballos_non_linear_MEMS_Robust} and will not be explained in further detail in the paper.

\subsection{GM-driven DFF-ROs}

 \begin{figure}[t]
     \centering
     \includegraphics[width=\linewidth]{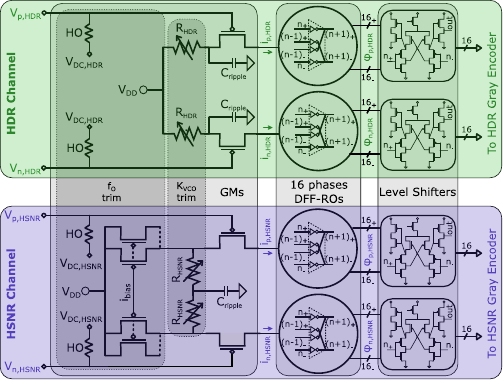}
     \caption{Schematic-level analog input stage of the Companding VCO-ADC.}
     \label{fig:full_analog}
 \end{figure}

  \begin{figure}[t]
     \centering
     \includegraphics[width=\linewidth]{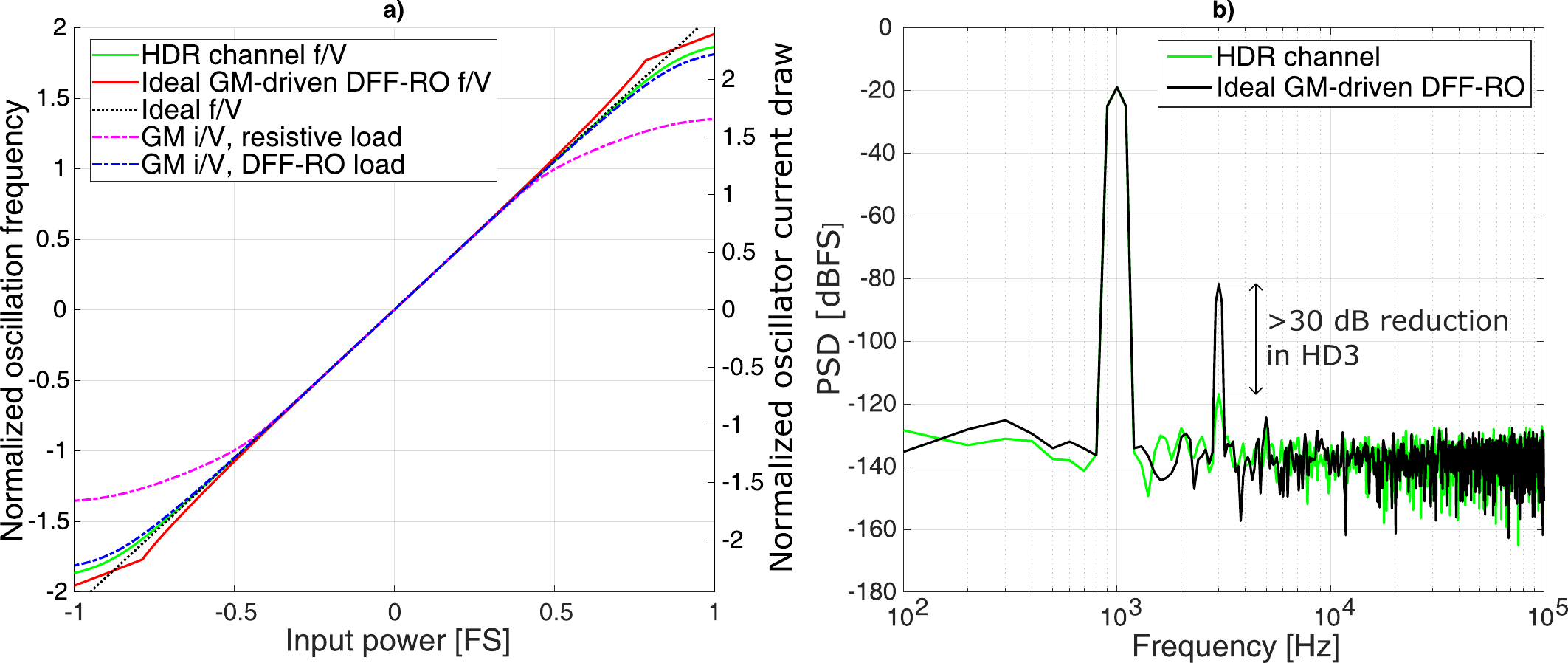}
     \caption{(a) Simulated HDR channel differential transfer curves.(b) Simulated PSD of the HDR channel at -13dBFS (without quantization noise), considering the implemented CCO driver and an ideal CCO driver.}
     \label{fig:hdr_sim_linearity}
 \end{figure}

  \begin{figure}[t]
     \centering
     \includegraphics[width=\linewidth]{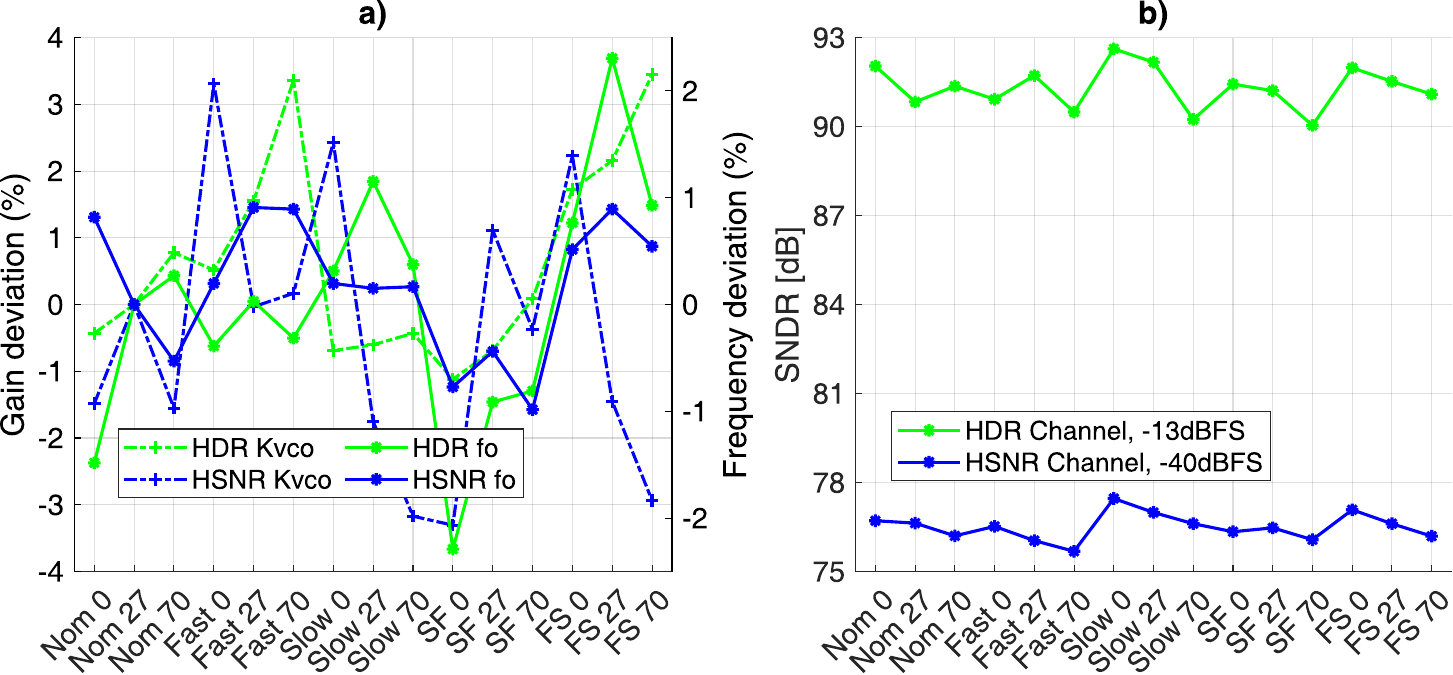}
     \caption{(a) Simulated frequency and gain drift across calibrated corners (b) Simulated SNDRs of HSNR and HDR channels in the calibrated corners.}
     \label{fig:Calibrated_corners}
 \end{figure}

The analog input stage of each channel of the implemented Companding VCO-ADC is shown in Fig. \ref{fig:full_analog}. Each channel includes a pseudo-differential source-degenerated transconductor driving the current-controlled oscillators. The HSNR channel transconductor is biased by a fixed current mirror, limiting the maximum current supplied to the oscillators to 2$i_{bias}$. This is done to ensure that, when the HSNR channel is saturated by a high-amplitude signal, no additional current is drawn by the ADC. On the other hand, the HDR channel requires higher linearity to achieve its intended DR. Additional voltage headroom in the GM+DFF-RO stage of the HDR channel is obtained by removing the current mirror that fixes the power draw of the oscillators. The DFF-RO oscillators are chosen because they can be configured to have a power-of-two stages, required for the phase-to-Gray encoder, and their IRN efficiency is superior to the conventional differential RO with direct cross-coupling \cite{borgmans_analog_VCO}. C$_{ripple}$ MOS capacitors (5pF) are placed in both channels to mitigate possible coupling across the $p$ and $n$ branches. After the RO, level-shifters based on \cite{lanuzza_level_shift} ensure the digital encoder logic works with adequate logic levels. Common-mode voltage to the branches is supplied through  high-ohmic (HO) resistors placed in parallel with the GM PMOS input, ensuring a high-impedance interface with the microphone MEMS. The Input Referred Noise (IRN) of the HSNR and HDR channels is 1.804$\mu$V$_{RMS}$ and 6.747$\mu$V$_{RMS}$ respectively, considering A-weighting of the noise floor and a bandwidth between 20Hz to 20kHz. \cite{Ceballos_companding}. The noise composition of the IRNs is shown in Table \ref{tab:noise_contributions}. In comparison with \cite{Carlos_DOC1}, the use of a lower effective oscillation frequency and a slightly higher oscillator area shifts the flicker noise contributions to the IRN to less than 25\% (with respect to 78.3\% in \cite{Carlos_DOC1}). This effect was previously mentioned in \cite{Medina_2nd}, although no noise contribution table was provided.

\begin{table}[t]
\centering
\scriptsize
\caption{A-weighted Noise Contributions}
\label{tab:noise_contributions}
\resizebox{\columnwidth}{!}{
\begin{tabular}{|l|c|c|c|c|}
\hline
\multirow{2}{*}{\textbf{Source}} & \multirow{2}{*}{\textbf{Device}}         & \multicolumn{1}{c|}{\multirow{2}{*}{\textbf{Noise Type}}} & \multicolumn{1}{c|}{\multirow{2}{*}{\begin{tabular}[c]{@{}c@{}}\textbf{Contribution in} \\ \textbf{HSNR Channel (\%)}\end{tabular}}} & \multicolumn{1}{c|}{\multirow{2}{*}{\begin{tabular}[c]{@{}c@{}}\textbf{Contribution in} \\ \textbf{HDR Channel (\%)}\end{tabular}}} \\
                        &                                 & \multicolumn{1}{c|}{}                            & \multicolumn{1}{c|}{}                                                                                              & \multicolumn{1}{c|}{}                                                                                             \\ \hline
\multirow{6}{*}{GM}     & \multirow{2}{*}{Current Mirror} & Thermal                                          & 26.6                                                                                                               & -                                                                                                                 \\ \cline{3-5} 
                        &                                 & Flicker                                          & 8.9                                                                                                                & -                                                                                                                 \\ \cline{2-5} 
                        & \multirow{2}{*}{Input PMOS}     & Thermal                                          & 6.0                                                                                                                & 3.1                                                                                                               \\ \cline{3-5} 
                        &                                 & Flicker                                          & 9.5                                                                                                                & 8.0                                                                                                               \\ \cline{2-5} 
                        & \multirow{2}{*}{Resistor}       & Thermal                                          & 17.6                                                                                                               & 27.2                                                                                                              \\ \cline{3-5} 
                        &                                 & Flicker                                          & 0.0                                                                                                                & 8.9                                                                                                               \\ \hline
\multirow{4}{*}{DFF-RO} & \multirow{2}{*}{PMOS}           & Thermal                                          & 13.4                                                                                                               & 24.3                                                                                                              \\ \cline{3-5} 
                        &                                 & Flicker                                          & 0.6                                                                                                                & 0.4                                                                                                               \\ \cline{2-5} 
                        & \multirow{2}{*}{NMOS}           & Thermal                                          & 12.9                                                                                                               & 20.8                                                                                                              \\ \cline{3-5} 
                        &                                 & Flicker                                          & 4.5                                                                                                                & 7.3     \\ \hline
\end{tabular}%
}
\end{table}

Looking at the HDR channel GM stage at high amplitudes (close to full-scale), the minimum current in an oscillator branch is limited by the voltage drop of $R_{HDR}$ reaching 0 (and thus, $f_o\rightarrow 0$). At the same time, the maximum current of an oscillator branch is also limited by the input PMOS entering the triode region, as the voltage drop in both the $R_{HDR}$ and the DFF-RO limits the operating range. These limits on the current draw of the HDR Channel DFF-RO may appear detrimental to the linearity of the input-voltage-to-frequency curve (f/V) of the GM+DFF-RO stage. Nevertheless, it is quite the contrary. Fig. \ref{fig:hdr_sim_linearity}(a) plots the differential transfer curves of the implemented HDR channel linearity-critical elements (GM and DFF-RO), normalized by the nominal current draw and rest oscillation frequency. The red line shows the DFF-RO oscillator being driven by an ideal current source, only accounting for the CCO nonlinearity. As the CCO reaches its lower frequency region, the f/i gain of the oscillators increases due to the oscillator being biased in weak inversion. At very high amplitudes, one of the DFF-ROs stops, producing a sharp gain difference in the differential curve. On the other hand, the i/V curve of the proposed GM stage (pink line) suffers poor linearity when driving a resistive load at the PMOS drain, due to the PMOS entering the triode region. However, when the GM drives the DFF-RO (blue line), the impedance of the oscillators drops as the current draw increases, resulting in a diode-like biasing behavior \cite{borgmans_analog_VCO}, and improving the SDR of the GM. In addition, the i/V gain of the GM tapers as the current decreases due to the GM PMOS entering its weak inversion region. This means that the DFF-RO f/i curve and the GM i/V curve can be co-designed to optimize the linearity of the resulting GM+DFF-RO f/V curve, and thus of the HDR channel. This is shown in Fig. \ref{fig:hdr_sim_linearity}(a) (green plot). This architecture choice results in a \textgreater 30dB HD3 improvement at the HDR channel peak SNDR, as shown in Fig. \ref{fig:hdr_sim_linearity}(b).

One drawback of open-loop VCO-ADC designs is their process and temperature sensitivity due to the lack of a closed feedback. Fig. \ref{fig:Calibrated_corners}(a) shows the gain and frequency calibrated corners against process and temperature (0-70ºC) of the HSNR and HDR channels. In both channels, gain is calibrated by trimming the GM 4-bit polysilicon resistor. The resolution of the HDR programmable resistor allows 0.4dB gain steps, resulting in a maximum 0.2dB analog gain difference between the channels. Regarding oscillation frequency, the HSNR channel uses a 4-bit programmable IDAC, biasing the oscillators (see Fig. \ref{fig:full_analog}), and the HDR channel uses an external 4-bit voltage DAC with an LSB of 10mV. As seen in the plots, the sensitivity drift of the channels is below $\pm$3.5\% ($\pm$0.3dB), and the rest oscillation frequency does not drift more than $\pm$2.5\% across corners. The noise of the HDR channel voltage DAC would not be critical, as it is common mode coupled and low-pass filtered by the HO resistors. In addition, both the HSNR and HDR channels are directly compatible with temperature-compensation schemes based on frequency-dependent CMFB \cite{DOG6_CMFB_FDR}. Regarding SNDR stability across corners, Fig \ref{fig:Calibrated_corners}(b) shows the HSNR Channel SNDR at -40dBFS (equivalent to 94dBSPL, see section V), as well as the simulated peak SNDR of the HDR channel at -13dBFS. As seen in the plots of  Fig. \ref{fig:Calibrated_corners}, the HSNR channel presents a maximum SNDR deviation of $\pm$1dB at -40dBFS, with deviations mainly related to increases in thermal noise contributions to the IRN. In the case of the HDR channel, all corners present a \textgreater 90dB peak SNDR, corroborating that the implemented GM+DFF-RO structure is capable of maintaining high linearity independently of process and temperature variations.

  \begin{figure}[t]
     \centering
     \includegraphics[width=0.85\linewidth]{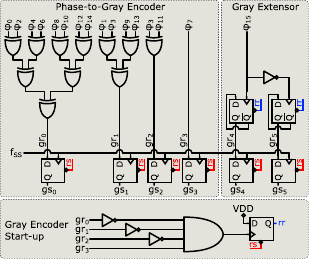}
     \caption{Schematic-level phase-to-Gray encoder.}
     \label{fig:gray}
 \end{figure}

 \subsection{Phase-to-Gray encoder}

Fig. \ref{fig:gray} displays the implemented phase-to-Gray encoder used both in the HDR and HSNR channels. Similar structures have been previously published by our research group \cite{Medina_2nd,DOGX_ESSERC}, but the authors would like to note that the implemented readout is sligthly more efficient (-2 flip-flops) than \cite{Medina_2nd} for the same input and output bus width. As explained in \cite{Medina_2nd,medinaGrayEncodedRingOscillator2023}, the proposed phase-to-Gray encoder allows a reduction in the number of flip-flops required to register the oscillator state (inverter-in-transition) from N to log$_2$(N)+1, but at the cost of adding N-(log$_2$(N)+1) XOR gates, being N the number of phases of the oscillator. This reduces the area and power requirements of the readout, effectively converting flip-flops to XOR gates, but requires the use of a differential ring oscillator with power-of-two phases (N=16 in the implemented case). Nevertheless, as the phases of the oscillator are directly encoded to a log$_2$(N)+1 bit Gray-coded signal, the circuit retains the maximum 1 LSB metastability error when sampled, as would directly sampling the oscillator phases. In the implemented design, the oscillators of both the HSNR and HDR channels have a rest oscillation frequency of $\approx$ 6MHz,  and the sampling frequency (f$_{ss}$) given by the delay line clock generator is $\approx$ 24MHz. This would mean that the oscillator state can be correctly sampled and processed by the synthesized digital datapath as f$_{ss}$\textgreater 2f$_{osc,max}$= 4f$_o$. However, as the f$_{ss}$ signal of the clock generator can present (or be configured) to have a non-uniform pulse pattern (see Fig. \ref{fig:XOR_DLL}), it would be possible that the digital datapath misses the RO making a complete oscillation cycle. This would reduce the maximum dynamic range of the channels, and make it dependent on the space between f$_{ss}$ pulses. As a solution, the implemented design uses a 1-bit Gray extensor, making the readout capable of encoding up to two complete oscillation cycles. Note that this extension is only 1-bit wide, as otherwise it would be possible to register directly $\varphi_{15}$ as gs$_4$ and obtain the 5-bit Gray-coded state of the oscillator. We include a circuit that resets the states of the gr$_4$ and gr$_5$ at start-up.

   \begin{figure}[t]
     \centering
     \includegraphics[width=0.85\linewidth]{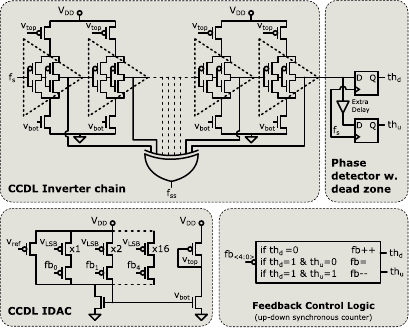}
     \caption{Schematic-level DLL-based unconstrained clock multiplier with deadzone.}
     \label{fig:dll_sch}
 \end{figure}

 \subsection{Delay-line based clock multiplier}

\begin{figure*}[t]
     \centering
     \includegraphics[width=0.95\linewidth]{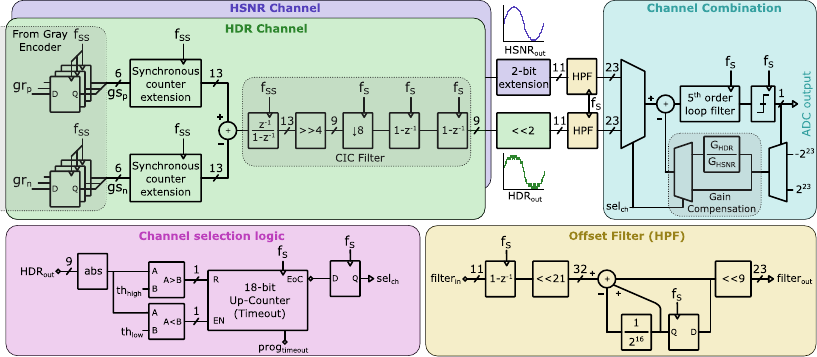}
     \caption{Companding VCO-ADC digital synthesized data-path.}
     \label{fig:digital_datapath}
 \end{figure*}

The clock multiplier block is built by means of a delay line, loosely controlled by a Delay Locked Loop (DLL). The working principle is based on a multiplication of the input master clock (f$_s$) to a pulsed signal with a higher frequency (f$_{ss}$). Fig. \ref{fig:dll_sch} illustrates a simplified circuit diagram of the proposed clock generator.

The core of the clock generator is a Current-Controlled Delay Line (CCDL). Its structure is comprised by a chain of current-starved inverters. The inverters are configured by connecting two in series as a buffer, and employing a total of 8 buffers (16 inverters). The output of each buffer is connected to a XOR gate, forming an XOR tree which outputs the desired multiplied clock. Note that the multiplication factor should correspond to the number of phases taken from the delay line. In this VCO-ADC architecture, the fast sampling rate is designed such that $f_{ss}=f_{s} \cdot M$ with $M=8$. Therefore, the number of phases captured by the XOR tree must be 8. The current mirror of the starved inverters in the CCDL is driven by 5-bit IDAC. The IDAC resolution limits the time resolution of delays $\tau_k$ on the CCDL. As a consequence, the total delay of the delay line may not match T exactly. 

However, in section III of the manuscript, it was explained that the present architecture using a CIC multi-rate filter is resilient to non-uniform sampling. In an extreme case, this would mean that it is possible to leave the delay line free-running without enabling the DLL feedback loop. In reality, the finite counter imposed by the phase-to-Gray encoder sets a limitation on the maximum gap that could be left between edges before reaching saturation. Hence, a control feedback logic is needed to ensure that this condition is met. Usage of a Bang-Bang phase detector would be a feasible feedback control loop. However, due to the limited resolution of the IDAC, the feedback control signal would oscillate at a frequency equal to f$_{s}$/2, resulting in a spur in the ADC output PSD. As an alternative, the control feedback loop is implemented using a phase detector with dead-zone and a simple up/down counter. In this case, a three state control logic is implemented by adding an small delay at the end of the delay line. The last phase of the chain and the extra phase are sampled, enabling the up/down counter to stay static while the phase shift stays between the sampled phases. The extra delay was designed to be approximately 2 times smaller than each inverter delay on the chain. As a summary, the DLL logic puts the delay line in the right bias condition at startup and then behaves as a watchdog function, checking that the delay does not get out of bounds (dead zone) due to temperature drifts.

\subsection{Synthesized digital datapath}
 
All the blocks described so far have been implemented using a custom analog layout. The rest of the ADC, explained in this section, has been synthesized and routed automatically using a standard logic library. A picture of the complete digital data-path is shown in Fig. \ref{fig:digital_datapath}. The first element of the data-path is a synchronous extension counter. The gray counter before sampling provides a total of 6 bits, which are not sufficient for the CIC filter operation that needs at least 13 bits \cite{laddomada}, and has been tested by simulation in our design. As a black box, the gray counter + binary conversion + synchronous extension behaves as a 13 bit counter that overflows periodically. To save power in the low-frequency side of the CIC filter, the 13 bits are truncated to 9 bits at the down-sampling stage. Given the block diagram of Fig. \ref{fig:CT_CIC_model}(a) (See Appendix A), any error introduced in the scaling factor between the last integrator and the first differentiator will be second-order noise-shaped, as it is succeeded by two first-order differences. This is the case for the truncation error (13 to 9 bits) that does not represent a limiting factor for the SNR.  Both the HSNR and HDR channel outputs are processed by identical CIC filters independently (blue and green blocks in Fig. \ref{fig:digital_datapath}), and thus require a multiplexer to combine them into a single ADC output. However, the signals are not ready for combination at this point, since both gain and offset calibration are needed.

A first-order high-pass filter is used to cancel the offset difference between channels. The filter (shown in Fig. \ref{fig:digital_datapath}, orange block) has a cutoff frequency around 7.5 Hz at a 3.072MHz sampling rate. Once filtered, both signals are effectively offset-free, requiring only gain mismatch compensation. At this point, a multiplexer is inserted to select one of the channels, leaving the gain calibration for the last stage. The channel selection logic (shown in Fig. \ref{fig:digital_datapath}, pink block) is similar to that in \cite{DOGX_ESSERC}, where it is extensively explained. In this ASIC implementation, the channel handover is detected within 5 T cycles. For a f$_s$ of 3.072MHz, this results in a latency $<2\mu$s. This ensures there are no clipping artifacts in the companding ADC output.

To keep compatibility with other MEMS digital microphones, the ASIC data output needs to be a (1 bit) PDM signal. A $5^{th}$ order digital noise-shaper is used for this purpose. Since the discrete-time sigma-delta modulators used for this purpose have a single bit feedback loop, this feedback can be exploited to match the different channel gain without using a hardware multiplier. The gain correction is accomplished by calculating the required gain difference between the analog HSNR and HDR channels, and modifying the feedback value corresponding to the full scale of the digital noise shaper accordingly. The noise shaper input values are 22 bit wide and the feedback must be two values corresponding to output symbols +1 or -1. A multiplexer is used to select the gain of the feedback, being 1 in case the HDR channel is selected, and a register with the computed value of G$_{HDR}$/G$_{HSNR}$ in the case of the HSNR channel is selected.

\section{Measurements}

\begin{figure}[t]
\centerline
\centering
\begin{center}
{\includegraphics [width=\columnwidth]{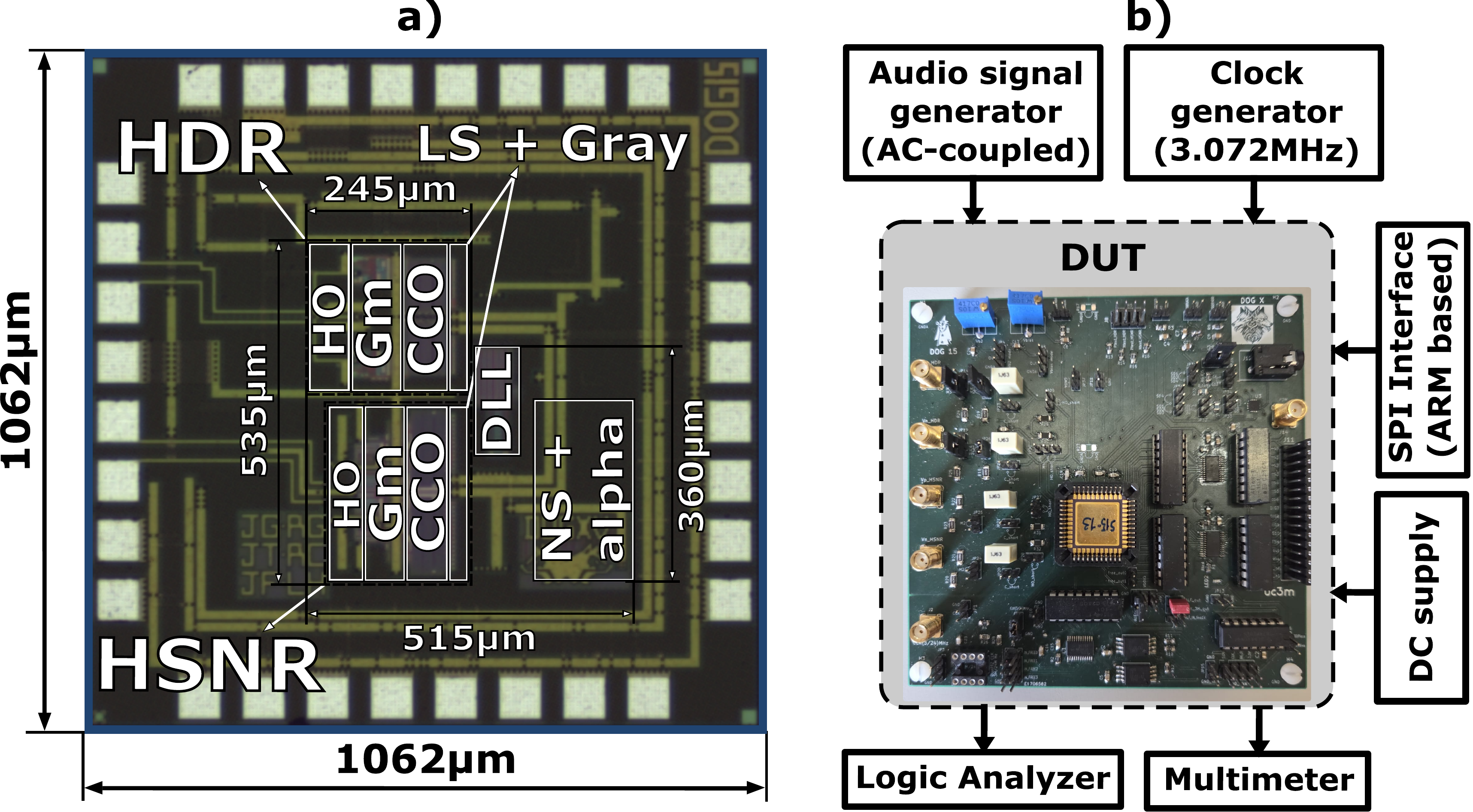}}
\end{center}
\caption{(a) Die micrograph.(b) Test setup. }
\label{fig:micrography_test_setup}
\end{figure}

A picture of the measurement setup, as well as the ASIC micrography, is shown in Fig. \ref{fig:micrography_test_setup}. Measurements were performed using low noise bench supplies, an audio and clock generator, a logic analyzer to download data and a micro-controller board as SPI programming interface. The total active area of the companding converter is 0.228 mm$^2$. A balanced differential audio signal is AC-coupled to both channels of the ADC, allowing that independent HO resistors provide the V$_{DC}$ bias voltage for each channel. The HDR channel employs a -6 dB internal capacitive attenuator before the GM gate ($C_{aten}$) \cite{Ceballos_companding}, ensuring that the input signal does not exceed the analog supply voltage at the gate. An additional -6dB attenuation is obtained thanks to the HDR Channel GM, resulting in the 12dB (4 times) analog gain difference between HSNR and HDR channels.

\begin{figure}[t]
\centerline
\centering
\begin{center}
{\includegraphics [width=\columnwidth]{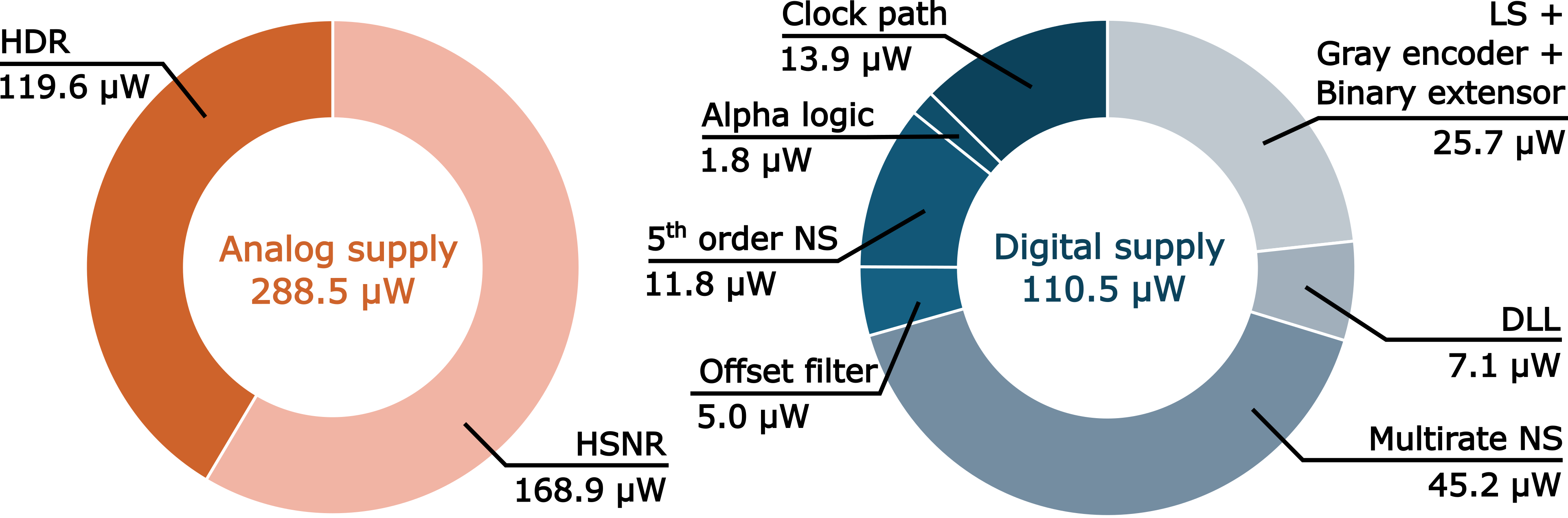}}
\end{center}
\caption{Measured power consumption breakout of the ADC, separating analog and digital supply domains.}
\label{power}
\end{figure}

A detailed measured power consumption chart of the ADC is shown in Fig. \ref{power}. The analog supply of the ADC is 1.5 V, while the digital supply is 0.9 V. The total power consumption is 399.0 $\mu$W. This power draw considers all functional elements of the ADC, the Clock generator with the DLL and the 1-bit-out noise-shaper, required for a MEMS microphone implementation. Single-channel operation modes are also available, with power consumptions of 187.0 $\mu$W if only the HDR channel is enabled, and 239.9 $\mu$W if only the HSNR channel is enabled. The total power consumption of the ASIC, including all auxiliary circuits (LDOs, band-gap) and the companding ADC core, is less than 500 µW. Despite the increased digital complexity of a companding ADC, power consumption is dominated by the analog supply. This is not only because of its higher V$_{DD}$ domain, but because the in-band noise of the ADC channels is dominated by thermal noise contributions (Table \ref{tab:noise_contributions}), and not by quantization noise. 

\begin{figure}[t]
\centerline
\centering
\begin{center}
{\includegraphics [width=0.9\columnwidth]{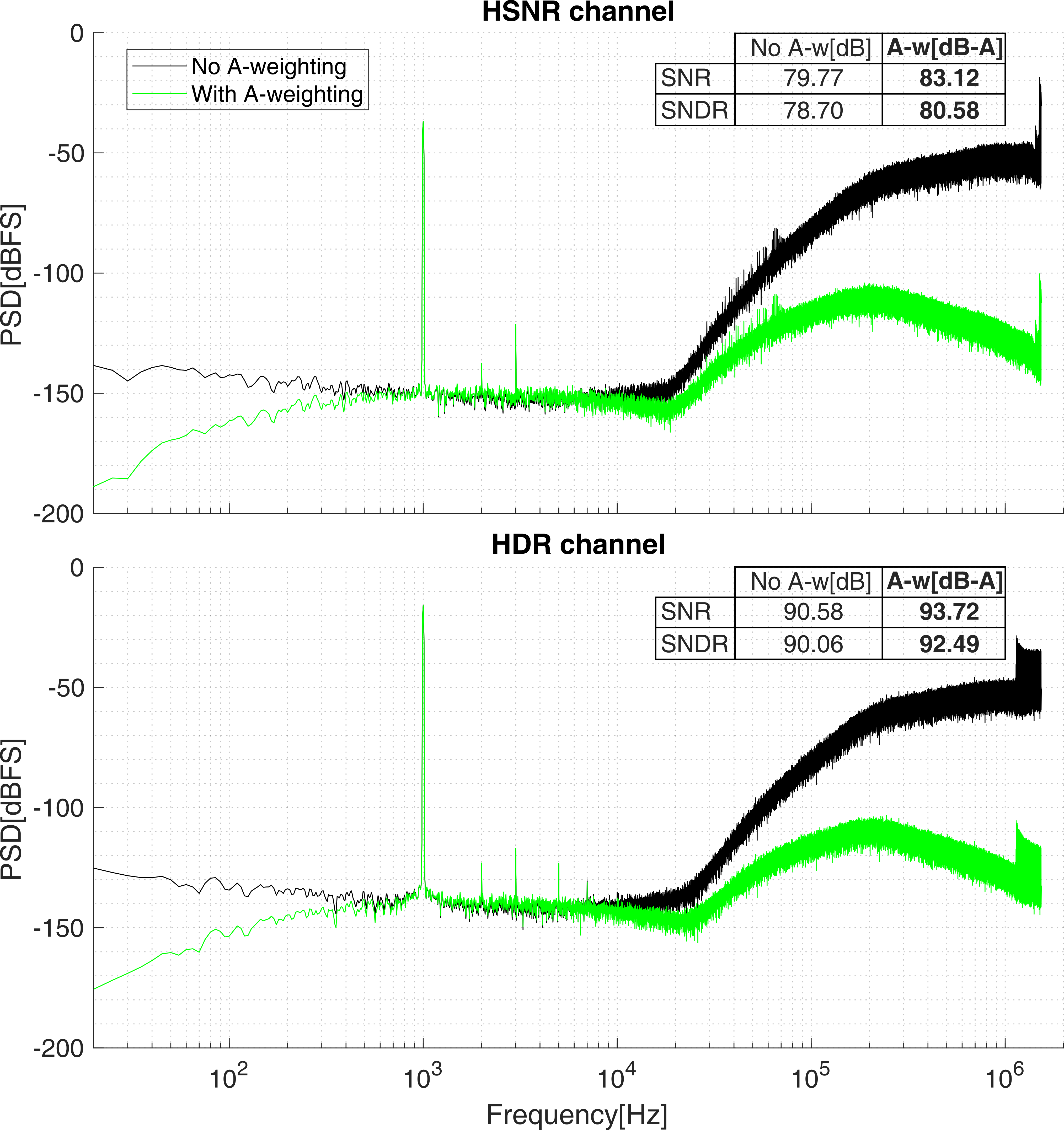}}
\end{center}
\caption{Measured PSD of peak SNDR of HDR and HSNR channels, at -34dBFS and -13dBFS respectively, single-bit output.}
\label{peak_SNDR}
\end{figure}

Fig. \ref{peak_SNDR} shows the measured PSD of peak SNDRs of the HSNR and HDR channels. These measurements are shown with and without A-weighting filter, as defined in DIN-IEC 651 \cite{a_weighting_Peus}. The A-weighting filter is standard in MEMS microphone ADC measurements, and allows fair comparison with leading industry works \cite{Ceballos_companding,Analog_amazing_Audio}. The A-weighting filter emphasizes middle tones (1 kHz - 6 kHz) and attenuating low and high tones. The peak SNDR of the HSNR channel is 78.7 dB, while the HDR channel achieves 90.1 dB. Linearity of the HSNR channel is penalized due to its GM structure, limiting the current that this stage draws. On the other hand, the GM structure and curve linearity optimization of the HDR channel proves effective to mitigate intrinsic CCO non-linearity. The measured peak SFDR of the HDR channel is 112.1 dBc at -22 dBFS. To the authors' knowledge, this is the highest SFDR measured in an open-loop VCO-ADC to date. 

\begin{figure}[t]
\centerline
\centering
\begin{center}
{\includegraphics [width=0.9\columnwidth]{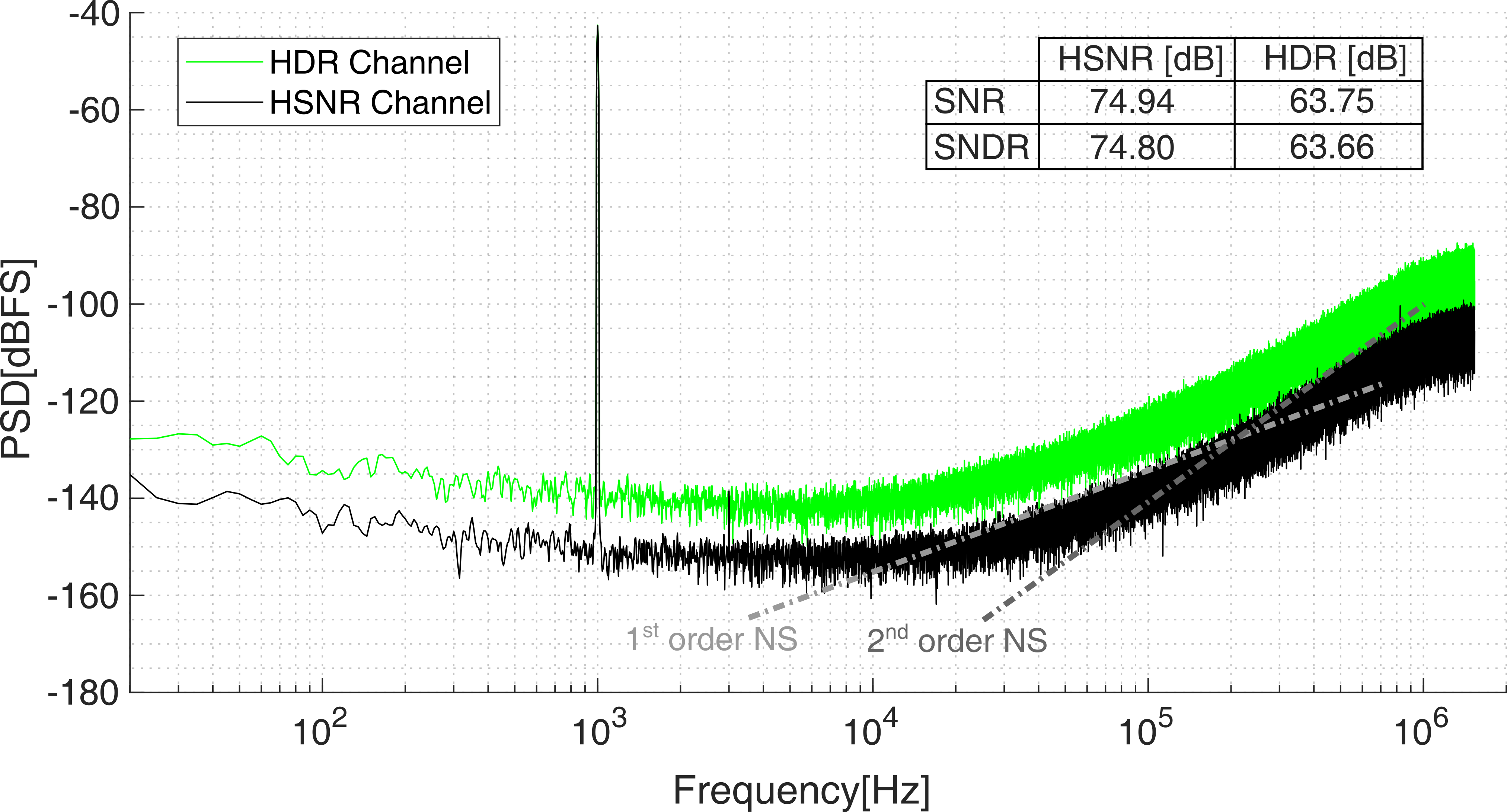}}
\end{center}
\caption{Measured PSD of HDR and HSNR channels without A-weighting at -40dBFS (94dBSPL), multi-bit output.}
\label{94dBSPL_SNDR}
\end{figure}

Regarding specifications of MEMS microphones, SNR of the ADC at 1 Pa of sound pressure (94 dBSPL) is an industry standard that will be used in our measurements. Fig. \ref{94dBSPL_SNDR} shows the measured PSD of the HDR and HSNR channels at 94 dBSPL, corresponding with -40 dBFS or an input voltage differential signal of -36 dBV considering a typical MEMS sensitivity \cite{Carlos_DOC1}. This PSD is obtained from the multi-bit multirate noise shaper, and clearly shows the first-order shaped noise, due to quantization noise, and second-order shaped noise due to the truncation from 13 to 9 bits which falls out of band (see Appendix A). Note the difference in the noise floor of the HSNR and HDR channels, due to their different input-referred noise targets (about 10 dB apart).  

\begin{figure}[t]
\centerline
\centering
\begin{center}
{\includegraphics [width=0.95\columnwidth]{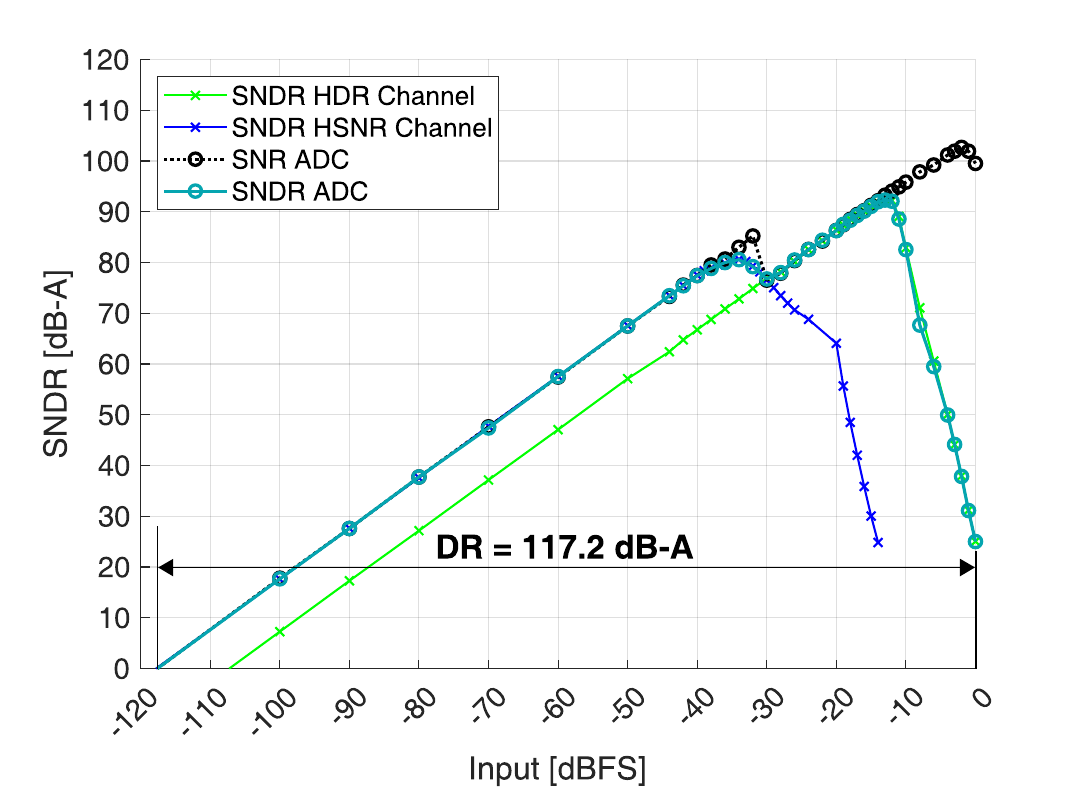}}
\end{center}
\caption{Measured DR of full ADC and HDR/HSNR channels, with A-weighting.}
\label{DR}
\end{figure}

The measured dynamic range of the companding ADC is shown in Fig. \ref{DR}. The switching point between the HSNR channel (for low amplitude input signals) and the HDR channel (for high amplitude input signals) can be programmed in the channel selection algorithm. In Fig. \ref{DR}, the switching point is configured at the crossing point between the dynamic ranges of the HSNR channel and the HDR channel (-30 dBFS). The reasoning behind this switch point is simple: when distortion starts being the main contributor to the THD+N in the companding ADC, it is desirable to commute to the higher linearity channel. When both channels are combined in a single digital output by means of the 1-bit noise shaper, the total dynamic range of the ADC reaches 117.2 dB-A. Note that, in audio, some linearity degradation at high amplitudes is tolerable. This is due to human hearing characteristics \cite{Carlos_DOC1} and also due to the linearity limitations of the capacitive MEMS connected to the input of the microphone \cite{ceballos_non_linear_MEMS_Robust}. We will define the Acoustic Overload Point of the ADC as the input level at which THD is 5\%, considering the nominal sensitivity \cite{Carlos_DOC1}. This more restrictive criteria (typically, AOP is defined at 10\% THD) is used as our measurements do not refer to a complete system with an acoustic-responsive MEMS. The dynamic range of the multirate noise-shaper has been designed to match with the companding ADC AOP to optimize area and power (see Fig. \ref{DR} SNR plot). Measured PSRR of the chip output across the audio bandwidth exceeds 100 dB in the HDR channel, and 110 dB in the HSNR channel, considering analog and digital supplies.

\begin{figure}[t]
\centerline
\centering
\begin{center}
{\includegraphics [width=0.9\columnwidth]{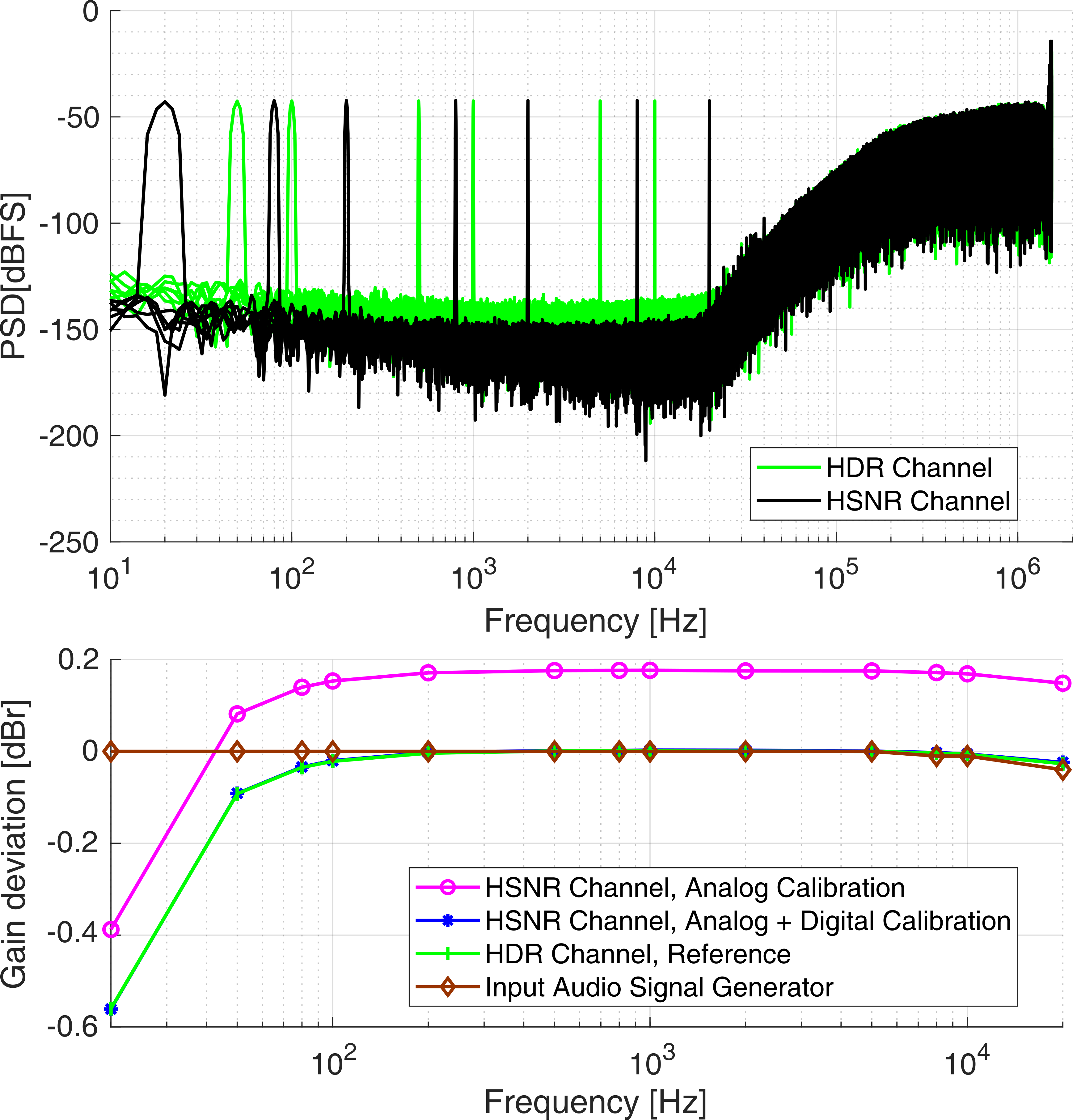}}
\end{center}
\caption{Measured frequency sweep of HSNR and HDR channels at -40dBFS, PSD and detailed view.}
\label{linearity}
\end{figure}

The intended companding ADC operation requires that both the HSNR and HDR channels present the same STF and signal gain at the channel output, ensuring that signals in the audio bandwidth do not present switching artifacts. Fig. \ref{linearity} plots a frequency tone sweep across the audio bandwidth of both channels, both as a PSD and as a signal gain deviation with respect to a 1 kHz reference tone. Checking the PSDs, no discernible spurs are detected at any tone frequency within the bandwidth. Checking the detailed signal gain deviation, it is shown that analog calibration does not completely match the gains of the HSNR channel and the HDR channel. This is due to the finite precision of the programmable R-DACs in the input GMs. A finer calibration of those gains is obtained digitally by the 1-bit noise shaper, as commented in section IV D and shown in Fig. \ref{linearity}. In addition, some signal attenuation at low and high frequencies is also seen. This attenuation is not because of the VCO-ADC STF, but because of the channel offset filter (at low frequency) and the input audio signal generator in the lab setup (at high frequency). 

\begin{figure}[t]
\centerline
\centering
\begin{center}
{\includegraphics [width=\columnwidth]{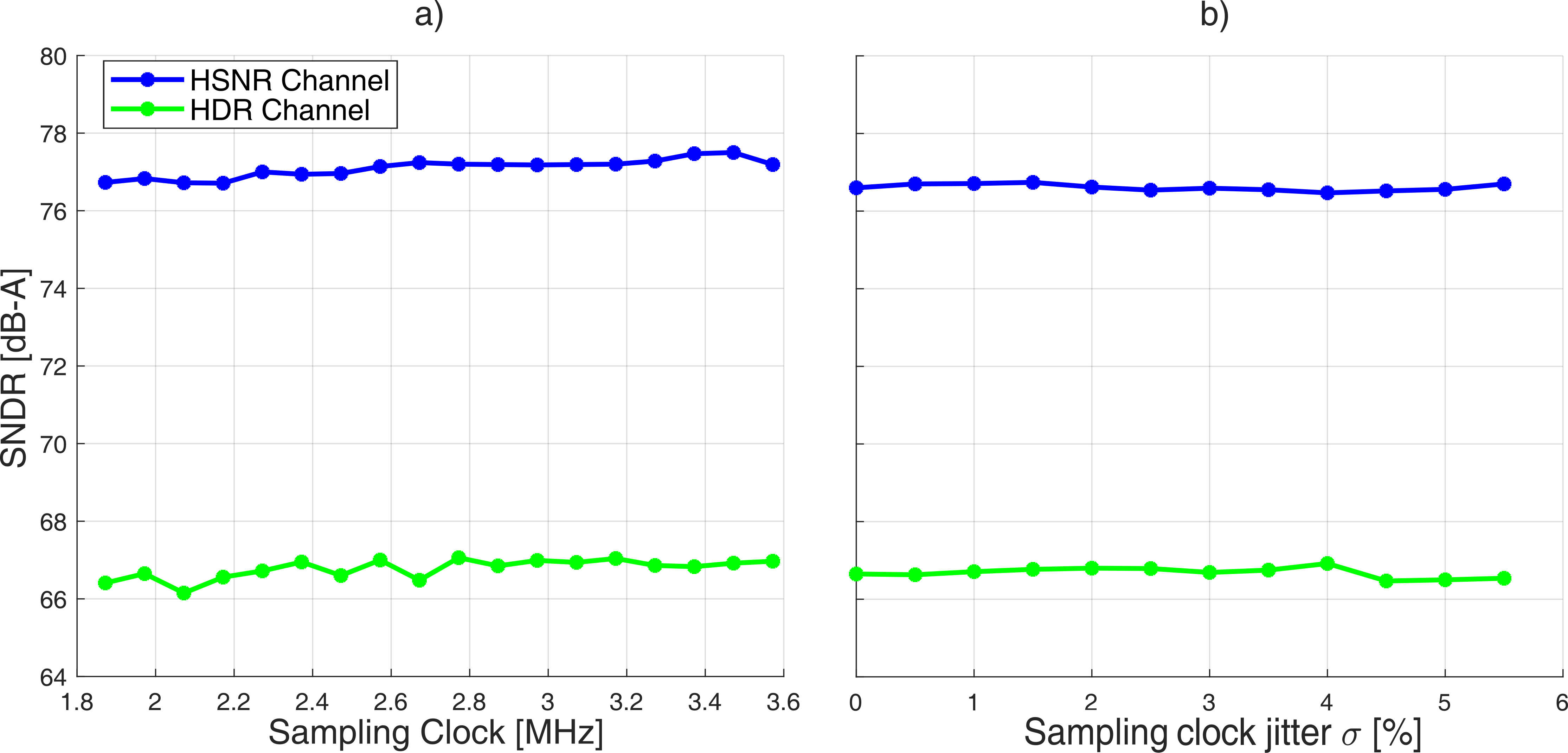}}
\end{center}
\caption{Measured SNDR at -40dBFS against sampling clock non-idealities:  (a) Sampling clock frequency sweep. (b) Sampling clock jitter sweep.}
\label{dll_fs_jitter}
\end{figure}

As explained in section III C, the clock generator based on a delay line (f$_{ss}$) allows a lower first-order quantization noise floor due to the up-sampling factor M. Nevertheless, its robustness against input clock non-idealities has not been demonstrated previously with measurements. Fig. \ref{dll_fs_jitter} shows measurements of each channel performance with respect to the input clock non-idealities. In these measurements, the delay line of the clock generator is calibrated once, then the feedback DAC code is held for the whole measurement sweep (DLL in open-loop). In addition, the measurements are performed at -40 dBFS, which corresponds to the noise-limited regime in both channels. In Fig. \ref{dll_fs_jitter}(a), a sampling clock frequency sweep is performed. Measurements show a maximum deviation in the SNDR (A-weighted) of 0.5 dB in both channels with respect to the unperturbed reference at 3.072 MHz. The reduced SNDR perturbation with respect to the input clock sweep is because both channels are noise-limited by thermal and flicker contributions, and not by quantization noise. On the other hand, the sampling clock may have an accurate central frequency, but suffer from clock jitter. Fig. \ref{dll_fs_jitter}(b) shows the effects of the sampling clock jitter in the channels SNDR. In this sweep, we also see a maximum 0.5 dB deviation with respect to the unperturbed input clock SNDR in all measurements of both channels, up to a jitter $\sigma$ of 5.5\%. 

\begin{figure}[t]
\centerline
\centering
\begin{center}
{\includegraphics [width=\columnwidth]{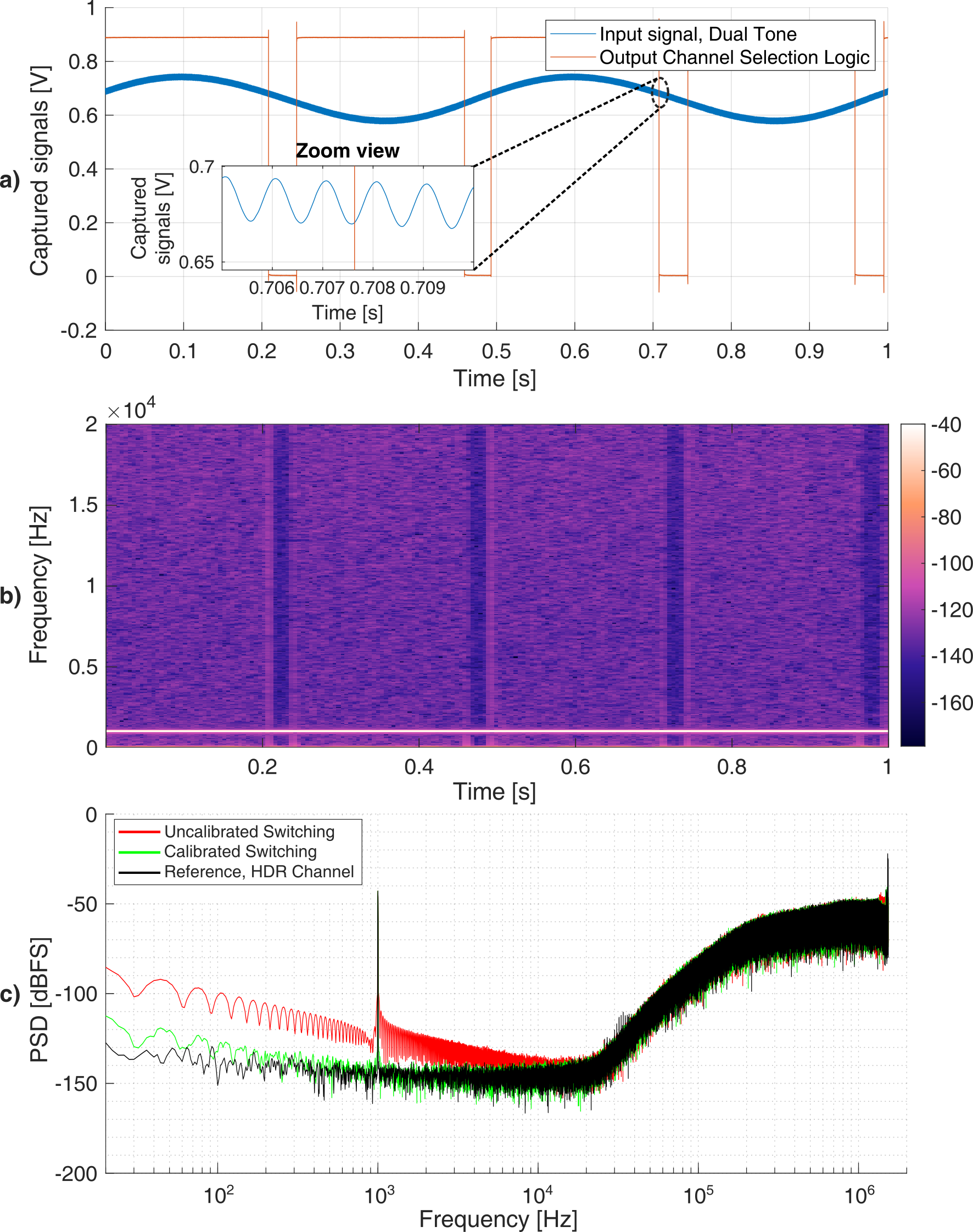}}
\end{center}
\caption{Measured switching performance: (a) Signal timeseries, (b) Extracted spectrogram of the on-chip calibrated 1-bit noise-shaped output (in dBFS) and (c) PSD comparison.}
\label{switching_performance}
\end{figure}

A deeper look into the audio companding ADC operation is given in Fig. \ref{switching_performance}. In this measurement, the companding ADC input signal is a dual-tone test input, with a subsonic component at 2 Hz and a sonic component at 1 kHz. The input power of the subsonic component is -20 dBFS, whereas the power of the sonic component is -40 dBFS. As the companding switching point is -30 dBFS, the subsonic component triggers the switching between the HDR channel and the HSNR channel periodically, being this component slow enough to exceed the timeout in the channel switch logic. This situation presents a worst-case scenario to the ADC due to the repeated channel switching. Fig. \ref{switching_performance}(a) shows the input dual-tone signal, as well as the output channel selection logic bit (1 for HDR channel, 0 for HSNR channel). In Fig. \ref{switching_performance}(b), the spectrogram of the on-chip calibrated 1-bit output signal of the ADC is shown. This spectrogram gives us information about when the HSNR and HDR channels are active at the ADC output, seen as differences in their noise floor power. In addition, this measurement also gives us information regarding transition glitches that occur when commuting between channels. Due to the nature of the digital gain calibration procedure through the 1-bit noise shaper, channel switching glitches are faintly appreciated in the spectrogram. A possible way to completely remove channel switching glitches is to perform the gain calibration not in the 1-bit noise shaper, but at the output of one of the channels (before the multiplexer). However, note that this fine gain calibration would require a multiplier in the channel signal path, where our solution does not. Furthermore, the authors think using the multiplier is an overkill solution, as the maximum perceived switching glitch in the spectrogram is 90 dB below the audible tone at 1 kHz. In a human audio test of the test input signal in Fig. \ref{switching_performance}, the switching glitch was not audible. To further reinforce this claim, an FFT of the output signal in Fig. \ref{switching_performance}(b) is also shown in Fig. \ref{switching_performance}(c). The PSD of the best-case scenario is shown in black, which is the output of only the HDR channel without the subsonic tone, and thus without channel switching. The line in green shows the PSD of the output signal used in Fig. \ref{switching_performance}(b). In this case, the SNDR only degrades by -0.4 dB-A. In case no digital calibration is used, the SNDR is heavily degraded by -16.7 dB-A. As a conclusion, the proposed calibration scheme is an effective solution in a companding VCO-ADC. 

\begin{table*}[t]
\centering
\caption{\label{table:comparison}Comparison with SOTA in audio ADC ASICs}
\begin{threeparttable}[b]
\resizebox{1.9\columnwidth}{!}{
\small\addtolength{\tabcolsep}{-4pt}
\def\arraystretch{1.5}

\begin{tabular}{l||c|c|c||c|c|c|c|c|c|c|c|c|c|c}

& \multicolumn{3}{c||}{\textbf{This work}} &  \cite{Huang_Mercier_VCO_audio} & \cite{Carlos_DOC1} & \cite{Medina_2nd}  & \cite{zhou_nansun_NestedCTDSM} &  \cite{Samsung_TWS} & \cite{Audio_P_martins}  & \cite{Analog_amazing_Audio} & \cite{Ceballos_companding} & \cite{Un_Ku_moon_PPD} & \cite{Ortmanns_audio_2025} & \cite{companding_100ms}\\

& Companding & HDR & HSNR & SSCL & IEEE Sensors  & TCAS I & CICC & ISSCC & CICC & ISSCC & ISCAS & ISSCC & JSSC & JSSC \\

& ADC & Channel & Channel & 2021 & 2023 & 2025 & 2024 & 2021 & 2025 & 2024 & 2024 & 2022 & 2025 & 2025\\ \hhline{=||=|=|=||=|=|=|=|=|=|=|=|=|=|=}

Process (nm) & \multicolumn{3}{c||}{130} & 65 & 130 & 130 & 180 & 28 & 65 & 130 & 130 & 180 & 180 & 180\\ \hline

Topology &  \multicolumn{3}{c||}{Open-loop} & Closed-loop & Open-loop & Open-loop  & Nested & & Zoom & & Companding & PPD & & Companding\\
& \multicolumn{3}{c||}{VCO-ADC} & VCO-ADC & VCO-ADC & VCO-ADC & CTDSM & CTDSM & CTDSM & CTDSM & DTDSM & DTDSM & DTDSM & DTDSM \\ \hline

Input & \multicolumn{3}{c||}{Hi-Z} & Low-Z & Hi-Z & Hi-Z & Low-Z & Low-Z & Low-Z & Low-Z & Hi-Z & Low-Z & Low-Z & Hi-Z\\ \hline

BW [kHz] & \multicolumn{3}{c||}{20} & 20 & 20 & 20 & 20 & 20 & 24 & 24 & 20 & 20 & 20 & 20 \\ \hline

Supply (A/D) [V] & \multicolumn{3}{c||}{1.5/0.9} & 1.0  & 1.5/0.95 & 1.5/0.95  & 1.8 & 1.8/1.1 & 1.2 & 3.3/1.8 & 1.8/0.9 & 1.8/1.1 & 1.8 & 1.8\\ \hline

F$_s$ [MS/s] & \multicolumn{3}{c||}{3.072} & 2  & 3.072 & 3.072 & 7.68 & 6.144 & 2.304 & 3.072 & 3 & 5.8 & 2.4 & 3.072\\ \hline

Area [mm$^2$] & \multicolumn{3}{c||}{0.228} & 0.11 & 0.14 & 0.095 & 0.74 & 0.07 & 0.36 & 0.48 & 1.1 & \textbf{0.0375} & 0.46 & 1.0\\ \hline

Power [$\mu W]$ & 399.0 & 187.0 & 239.9  & 142.6 & 438.1 & 250 & 470 & \textbf{116} & 208 & 3310 & 460 & 203.5 & 470 & 774\\ \hline

SNR$_{peak}$ [dB],[dB-A] & 100.9/103.2$^{\triangle}$ & 100.9/103.2$^{\triangle}$ & 99.6/102.3$^{\triangle}$ & 97.3 & & 97$^{\triangle\square}$ & & 100.7 & 105.7 & & 92$^{\triangle\square}$ & \textbf{106.7} & 102.2 & 68.2$^{\triangle}$$^{\star}$ \\ \hline

SNDR$_{peak}$ [dB],[dB-A] & 90.4/92.2$^{\triangle}$ & 90.4/92.2$^{\triangle}$ & 78.8/80.0$^{\triangle}$ & 94.2 & 80.3$^{\triangle}$ & 76.5$^{\triangle}$ & 107.3 & 100.6 & \textbf{105.4} & 103.9 & & \textbf{105.4} & 100.7 & 68.2$^{\triangle}$$^{\star}$ \\ \hline

SFDR$_{peak}$ [dBc] & 112.1 & 112.1 & 97.2 & 115.6 & &  & 118 & \textbf{126.5} & & & & 115 & \\ \hline

DR [dB],[dB-A] & 114.3/117.2$^{\triangle}$ & 104/106.8$^{\triangle}$ & 100.4/103.3$^{\triangle}$ & 100.3 & 108$^{\triangle}$ & 103$^{\triangle}$ & 109.2 & 104.4 & 106.3 & \textbf{116.3/118.5$^{\triangle}$} & 108$^{\triangle}$ & 108.8 & 102.6 & 107.2$^{\triangle}$$^{\star}$  \\ \hline

FoM$_{SNDR}$ [dB],[dB-A] & 167.4/169.2$^{\triangle}$ & 171.0/172.8$^{\triangle}$ & 158.2/159.3$^{\triangle}$ & 175.7 & 157$^{\triangle}$ & 155.5$^{\triangle}$ & 183.6 & 183.7 & \textbf{186.0} & 171.7 & & 185.3 & 178.9 & \\ \hline

FoM$_{DR}$ [dB],[dB-A] & \textbf{191.3/194.2$^{\triangle}$} & 184.3/187.1$^{\triangle}$ & 179.6/182.5$^{\triangle}$ & 181.8 & 184.6$^{\triangle}$ & 182$^{\triangle}$ & 185.5 & 187.5 & 186.9 & 184.1/186.3$^{\triangle}$ & 184.4$^{\triangle}$ & 188.7 & 177.0 & 181.3$^{\triangle}$$^{\star}$ \\

\end{tabular}}

\begin{tablenotes}
    \item $^{\triangle}$Use of A-weighting filter.    $^{\square}$Estimated from figures. $^{\star}$Acoustic measurements with a capacitive MEMS.
\end{tablenotes}
\end{threeparttable}
\end{table*}

Table \ref{table:comparison} compares the measured companding VCO-ADC architecture against other state-of-the-art solutions for audio ADCs. Our results are given with and without the use of the A-weighting filter for fair comparison. In addition, we provide results for the full companding ADC, as well as for each ADC channel independently. Both the reported SNDR and DR figures of our work are taken as a 5-sample chip average, being those results used for the FoMs. Compared with other open-loop VCO-ADCs \cite{Carlos_DOC1,Medina_2nd} and other works with Hi-Z input impedance \cite{Ceballos_companding}, our work boasts higher FoMs both in the companding configuration and using only the HDR channel. Regarding linearity, our work shows comparable SFDR with \cite{Huang_Mercier_VCO_audio}, which employs a closed-loop VCO-ADC topology and would require an audio input buffer for connection with a MEMS device. Area and power consumption of the ADC are comparable with state-of-the-art solutions using CTDSM and DTDSM, which are common in audio applications. FoM$_{SNDR}$ is significantly better in CTDSM and DTDSM than in our companding VCO-ADC thanks to the negative DAC feedback path. At the same time, both DTDSM and CTDSM architectures need an audio input buffer to be compatible with MEMS microphones because of this feedback path, which is not typically included in the FoM. Furthermore, the capacitive MEMS readout does not require such high linearity in the first place due to the MEMS intrinsic distortion at high amplitudes. As such, this work's main target is to have as high a dynamic range as possible within its target power consumption, under 400 $\mu$W. Our work boasts the highest FoM$_{DR}$ with respect to the compared solutions, considering the full companding ADC solution. In comparison with other companding architectures for MEMS microphones \cite{Ceballos_companding,companding_100ms}, our work achieves better performance and area thanks to the use of time-domain ADCs. In addition, it has the second-highest dynamic range, only surpassed by \cite{Analog_amazing_Audio}. Nevertheless, note that \cite{Analog_amazing_Audio} requires an almost order of magnitude higher power consumption to achieve its DR. As a side note, the HDR channel considered as a stand alone ADC shows a comparable FoM$_{DR}$ to the state-of-the-art solutions, reinforcing the proposed multi-rate sampling scheme as an efficient solution for VCO-ADCs.

\section{Conclusions}
As main conclusion of our paper, we have experimentally proven the viability and performance attainable with VCO-ADCs in digital MEMS microphones employing a companding architecture. We have also used this example as a vehicle to analyze and prove the operation of the multi-rate VCO-ADC architecture. The multi-rate VCO-ADC has shown a surprising tolerance to clock imperfections which has been thoughtfully analyzed form a theoretical point of view and verified experimentally. Comparing our work to previous audio ADCs and VCO-ADCs, it shows a state of the art FoM both from using the definitions based on DR and peak SNDR. As additional highlights of our chip, we have shown how by co-optimizing the driver and oscillator design, the linearity of the VCO-ADC can be in par with other ADC topologies, attaining 112.1dB of SFDR. This demystifies the idea that open-loop VCO-ADCs are not linear enough for practical applications.

\begin{appendices}
\section{System level analysis of the multi-rate VCO-ADC}

\begin{figure}[t]
\centering
{\includegraphics [width=\columnwidth]{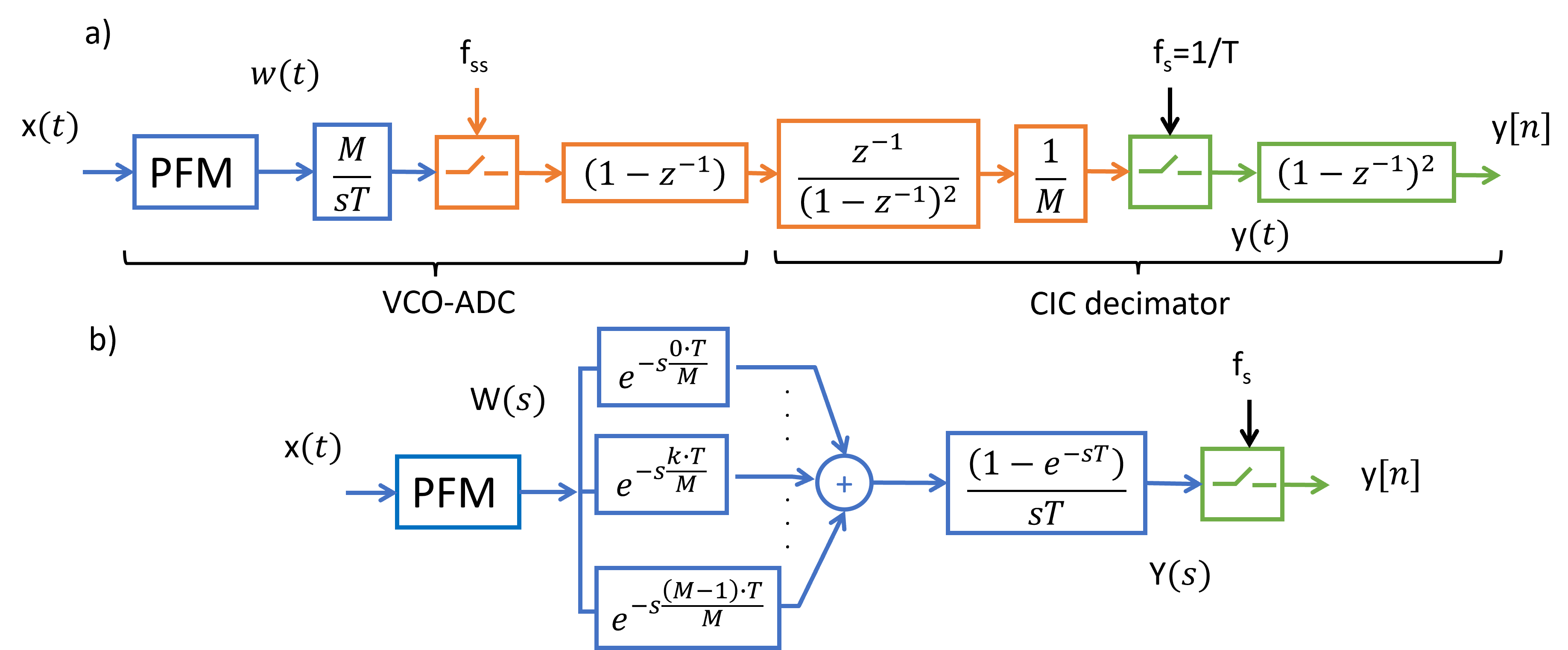}}
\caption{(a) Discrete-time block diagram of the multi-rate VCO-ADC (b) Continuous-time block diagram of the multi-rate VCO-ADC.} 
\label{fig:CT_CIC_model}
\end{figure}

\begin{figure}[t]
\centering
{\includegraphics [width=0.9\columnwidth]{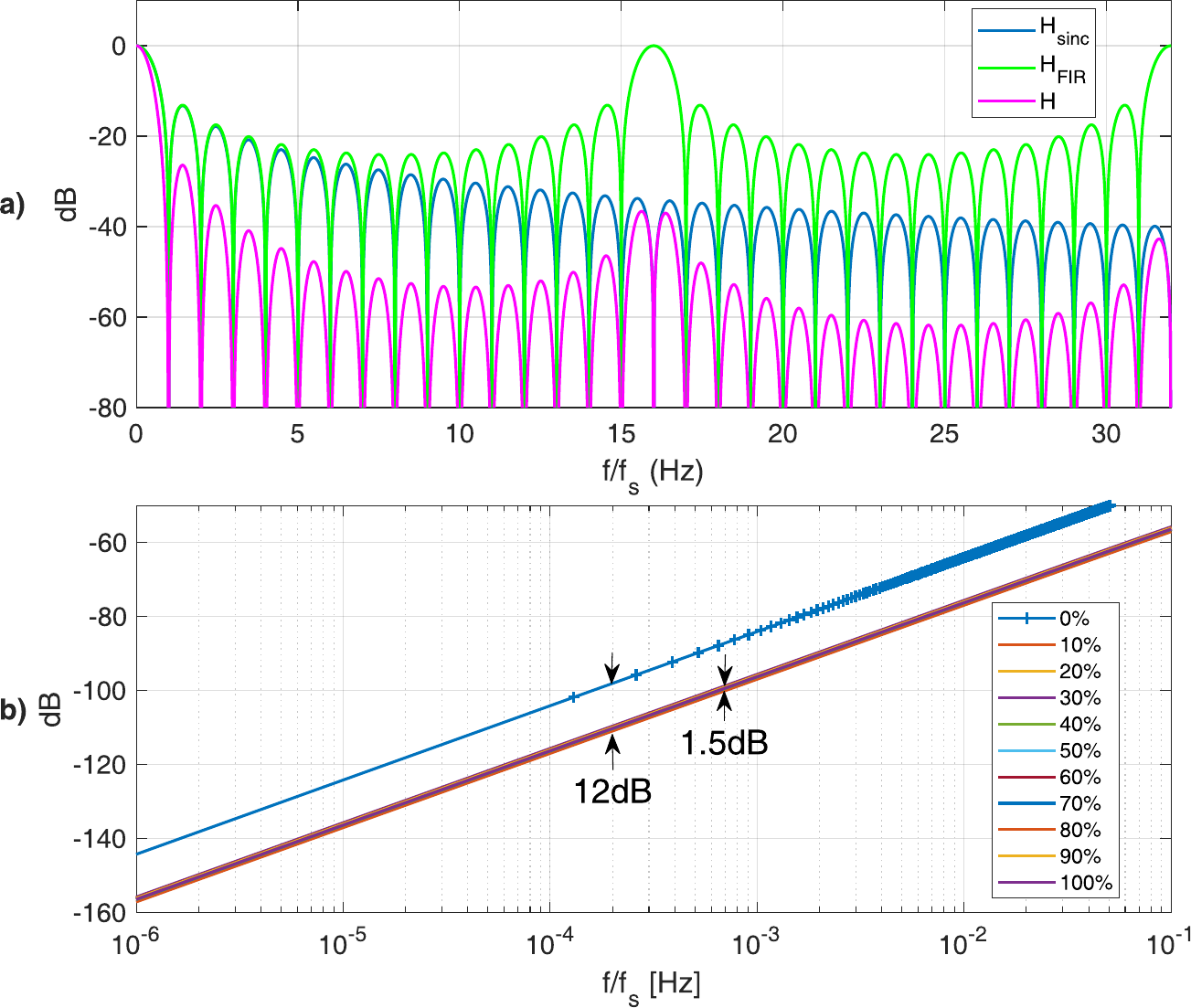}}
\caption{(a) Transfer functions in the multi-rate VCO-ADC (b) Aliased transfer function under delay mismatch.} 
\label{fig:tfs}
\end{figure}

\begin{figure}[t]
\centering
{\includegraphics [width=0.9\columnwidth]{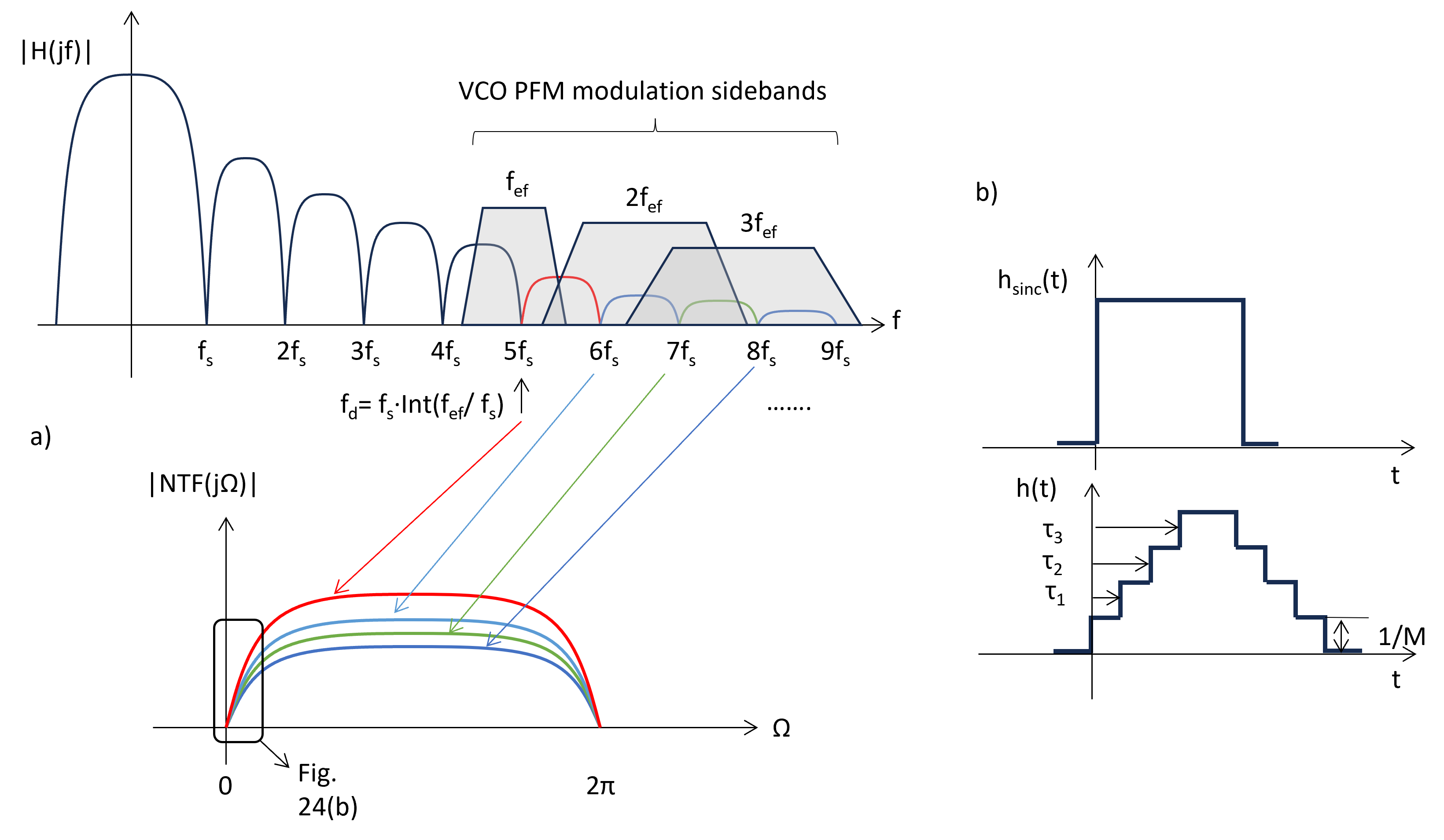}}
\caption{(a) Quantization noise production mechanism in a VCO-ADC. (b) Continuous-time impulse response of multi-rate VCO-ADC filter} 
\label{fig:folding}
\end{figure}

The minimal influence of timing errors in the SQNR of the multi-rate frequency-to-digital converter of Fig. \ref{fig:CF_vs_DLLCIF}(b), has motivated the analysis of this subsection. The analysis is based on the Sigma-Delta modulation theory using Pulse Frequency Modulation (PFM), described in \cite{medinaDifferentViewSigmaDelta2024}. We will use the VCO-ADC model of \cite{hernandezAnalyticalEvaluationVCOADC2015} as it will enable us to represent the VCO-ADC under non-uniform sampling conditions. We will assume first that sampling clocks $f_{ss}$ and $f_{s}$ are uniform. This model is displayed in Fig. \ref{fig:CT_CIC_model}(a), where the VCO is represented by a Pulse Frequency Modulator (PFM) block that encodes input $x(t)$ into a Dirac delta pulse train $w(t)$. The PFM signal is integrated by the RO inherent counting function \cite{magazine_colorines_pt1}. Afterwards, we have placed the sampler at frequency $f_{ss}$ followed by a first difference to complete the VCO-ADC. We have also detailed the subsequent discrete-time blocks of the CIC filter of Fig. \ref{fig:CF_vs_DLLCIF}(b). To implement the $2^{nd}$ order CIC filter, we have two integrators at $f_{ss}$, a decimation by factor M and two first-difference functions at rate $f_{s}$. To clarify Fig. \ref{fig:CT_CIC_model}, we have depicted in blue the continuous time blocks, in red those operating at $f_{ss}$ and in green those operating at $f_{s}$. We will analyze an alternative representation of Fig. \ref{fig:CT_CIC_model}(a)  obtained by moving all the discrete time delays $z^{-1}$ to the continuous time domain. By replacing the unit delays operating at $f_{s}$ (green) by $e^{-sT}$ and those operating at $f_{ss}$ (red) by $e^{-sT/M}$, we may use a single sampler at the end controlled by clock $f_{s}$. The continuous-time transfer function is:

\begin{eqnarray}
H(s)=Y(s)/W(s)=\cfrac{e^{-sT/M} \cdot (1-e^{-sT})^2}{sT \cdot (1-e^{-sT/M})}
\label{eq:laplace_y}
\end{eqnarray}

We can also express \eqref{eq:laplace_y} using a poly-phase decomposition with M interleaved channels sampled at $f_{s}$. This modification, reveals that the CIC decimator can be decomposed into a continuous-time FIR filter and a sinc filter after the FIR filter with a sampler at the end (see Fig. \ref{fig:CT_CIC_model}(b)). The delays of the FIR filter match with the delays of the delay line of Fig. \ref{fig:XOR_DLL}(b):

\begin{eqnarray}
H(s) = H_{FIR}(s) \cdot H_{sinc}(s)  \\
H_{FIR}(s)= \cfrac{1}{M} \cdot \sum^{M-1}_{k=0} e^{-s\tau_k}    , H_{sinc}(s) = \cfrac{(1-e^{-sT})}{sT} 
\label{eq:laplace_yfir}
\end{eqnarray}

 In the uniform sampling case, delays $\tau_k$ in \eqref{eq:laplace_yfir} will all be $\tau_k=k \cdot T / M$. Fig. \ref{fig:tfs}(a) represents the modulus of several transfer functions in \eqref{eq:laplace_yfir}: $H_{FIR}(j \omega)$ in green,  $H_{sinc}(j \omega)$ in blue and the product of both functions $H(j \omega)$ in pink. All functions are plotted normalized to frequency $f_s$ with $M=16$. Function $H(s)$ shows some periodic zeros which are double (second order) except every M zeros where they are single (first order). As a consequence, the aliased components from the VCO modulation sidebands \cite{gutierrezPulseFrequencyModulation2018, hernandezAnalyticalEvaluationVCOADC2015} will be dominated by the first order nulls and produce first order noise shaping but with smaller noise than if we were sampling the VCO at $f_s$. This extra SNR comes form the FIR attenuation over the sinc and corresponds to factor $10 \cdot log_{10}(M)$ in \eqref{eq:dr_DLL_cif}. We can approximate the Noise Transfer Function (NTF) of the VCO-ADC by the combination of the infinite aliases of $H(s)$ sampled at $f_s$. In our approximation, we are interested in estimating the SQNR of moderate or small signals, before SNDR is limited by distortion. Therefore, we have restricted our alias calculation to the frequencies where the Pulse Frequency Modulation (PFM) sidebands \cite{gutierrezPulseFrequencyModulation2018} produced by the VCO are centered, which are all frequencies above $f_d$, the multiply of $f_s$ closer to $f_{ef}$. This situation is depicted in Fig. \ref{fig:folding}(a) where we have sketched H(s) and the PFM modulation sidebands. An approximation of the modulus of the NTF is:

\begin{eqnarray}
f_d=f_s \cdot Int [f_{ef}/f_s]  , 0 \leq \Omega< 2\pi \nonumber \\
|NTF(j\Omega)|^2 \approx \sum_{i=0}^{\infty} | H(2\pi j (f_d+i \cdot f_s)+j \Omega f_s)|^2   
\label{eq:ntf_mod}
\end{eqnarray}
 
  The continuous-time representation of the VCO-ADC, \eqref{eq:laplace_yfir} is still valid if we assume that clock $f_{ss}$ is periodically irregular and defined by the delays of the delay chain of Fig. \ref{fig:XOR_DLL}(b), as long as the sampler at $f_s$ remains at the end. To test the sensitivity to the timing mismatch, we have calculated the modulus of $NTF(j\Omega)$ (see \eqref{eq:ntf_mod}) in the simulation conditions of Fig. \ref{fig:sim_gap} when all delays $\tau_k$ are equal but shrunk between $100\%$ up to $10\%$ of their nominal value $kT/M$, in $10\%$ steps. Note that the extreme case $\tau_k=0\%$ of $kT/M$ corresponds to the case $f_{ss}=f_s$. We have plotted in log scale the NTF of \eqref{eq:ntf_mod} in the analog bandwidth defined by OSR=77 up to $i=10^3$. The plot is shown in Fig. \ref{fig:tfs}(b), where we can see the 20dB/dec roll-off corresponding to $1^{st}$ order noise shaping. Except for $f_s=f_{ss}$, all plots fall within 1.5dB of the ideal case where $\tau_k=kT/M$. Also, the difference between the extreme case $M=1$ and the nominal one $f_{ss}=M \cdot f_s$, matches exactly with 12dB, the expected value according to \eqref{eq:dr_DLL_cif}.   
  
  An intuitive explanation on why there is an abrupt change in the SQNR in Fig. \ref{fig:tfs}(b) can be found in the unit impulse response $h(t)$ corresponding to transfer function $H(s)$ (see Fig. \ref{fig:folding}(b)). When delays $\tau_k$ are nonzero, $h(t)$ resembles a pyramid with M steps up and then down after T seconds. This translates in M times more quantization levels at the output of the modulator, therefore more SQNR. However, when all delays are equal, ($f_s=f_{ss}$), the impulse response turns abruptly into a square pulse as in $h_{sinc}(t)$ with a single level. Note that the steps in $h(t)$ are sized 1/M regardless of the timing mismatch, therefore the timing mismatch does not incur in a nonlinear effect but in a different attenuation of the FIR filter.  

\end{appendices}

\bibliography{references}
\bibliographystyle{IEEEtran}

\begin{IEEEbiography}[{\includegraphics[width=1in,height=1.25in,clip,keepaspectratio]{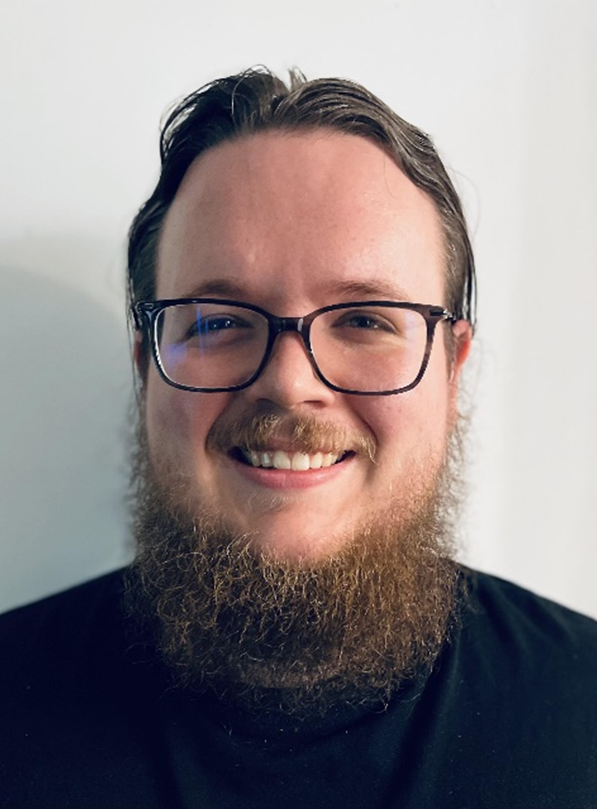}}] {Javier Granizo} received the M.E. degree in electronic engineering from Carlos III University of Madrid in 2021, where he is currently pursuing a Ph.D. degree. In 2024, he has done a three month stay at Arizona State University (ASU). His current research interests include VCO-ADCs, spiking neural networks, and bio-signal processing. 
\end{IEEEbiography}

\begin{IEEEbiography}[{\includegraphics[width=1in,height=1.25in,clip,keepaspectratio]{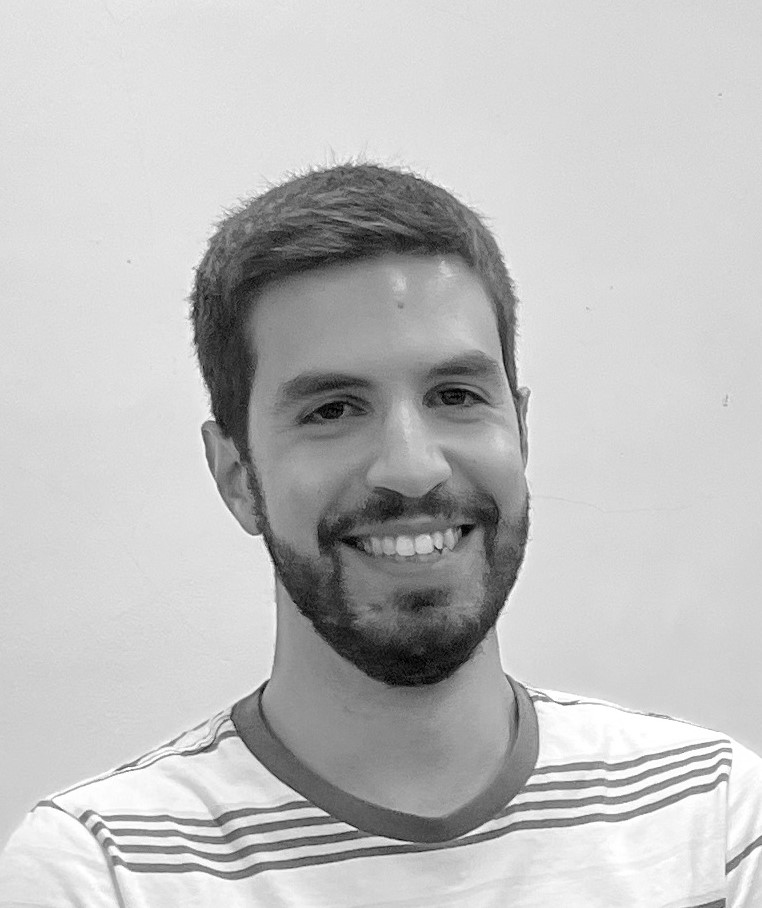}}] {Ruben Garvi Jimenez-Ortiz} received the B.Sc., M.E. and PhD degrees in Electronic engineering from Carlos III University, Madrid, Spain, in 2016, 2017 and 2023 respectively. In 2019 he did a four month internship at Infineon Technologies Austria. His current research interests include mixed-signal integrated circuits design and MEMS sensors. He is currently an Assistant Professor at Carlos III University. 
\end{IEEEbiography}

\begin{IEEEbiography}[{\includegraphics[width=1in,height=1.25in,clip,keepaspectratio]{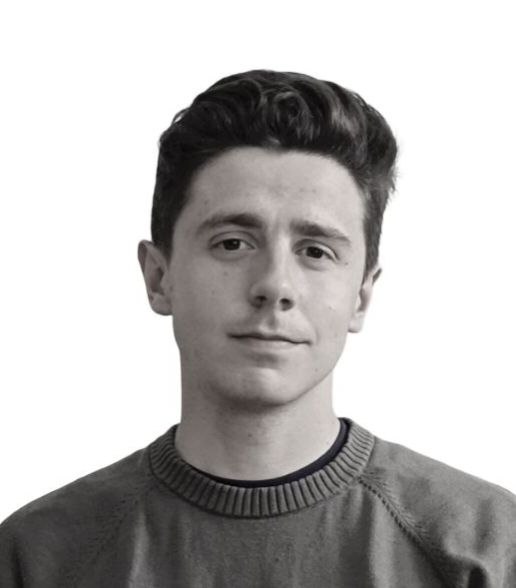}}] {Ricardo Carrero} received the B.Sc. degree form Polytechnic University of Madrid and the Ms degree in Electronic engineering from Carlos III University, Madrid, Spain, in 2022 and 2024 respectively, where he cooperated in the development of this work. His current interests are mixed-signal integrated circuit design and digital design. 
\end{IEEEbiography}

\begin{IEEEbiography}[{\includegraphics[width=1in,height=1.25in,clip,keepaspectratio]{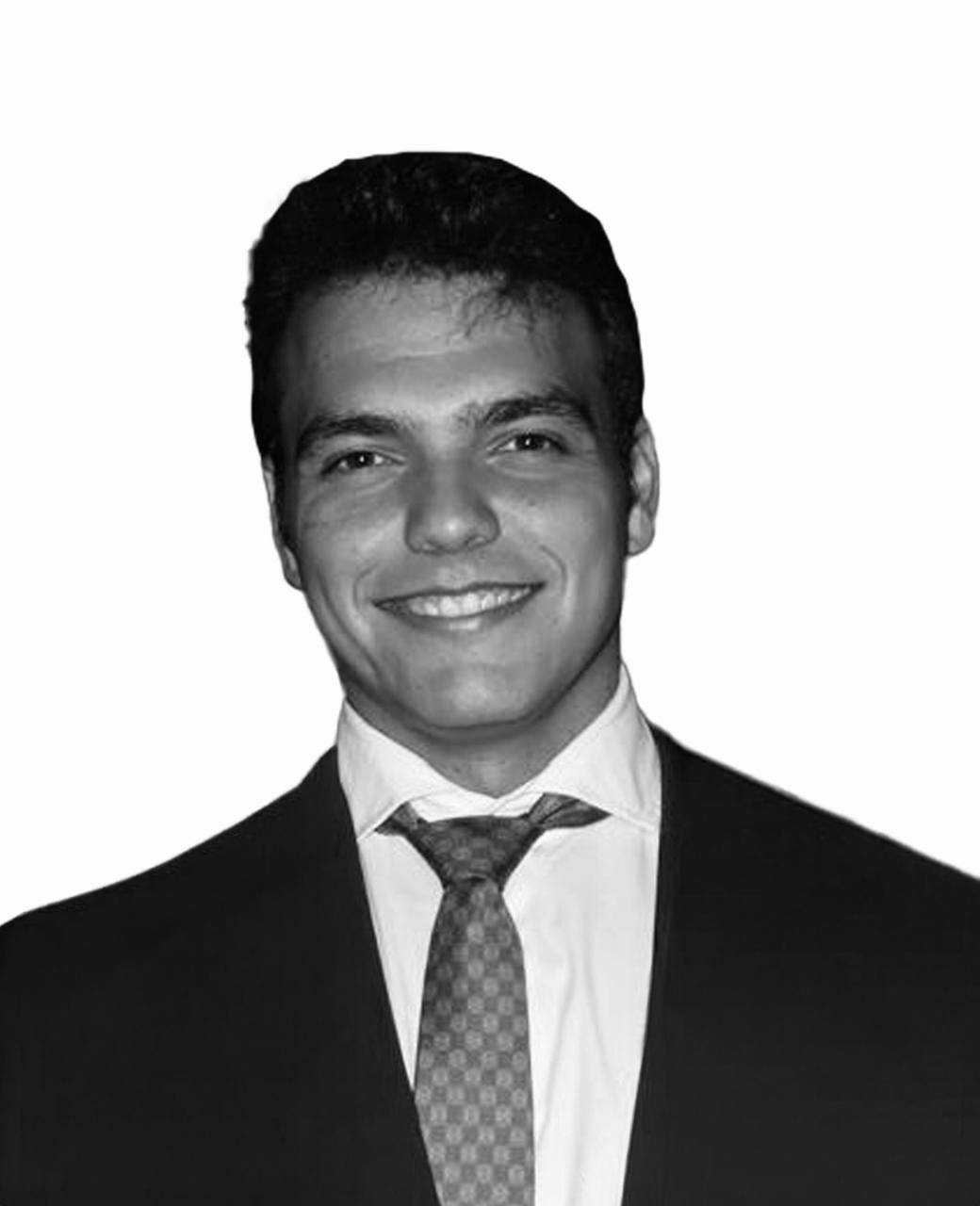}}] {Jorge de la Torre} received the B.Sc. and the Ms degrees in Electronic engineering from Carlos III University, Madrid, Spain, in 2023 and 2024 respectively, where he cooperated in the development of this work. His current interests are mixed-signal integrated circuit design and data converters.
\end{IEEEbiography}

\begin{IEEEbiography}[{\includegraphics[width=1in,height=1.25in,clip,keepaspectratio]{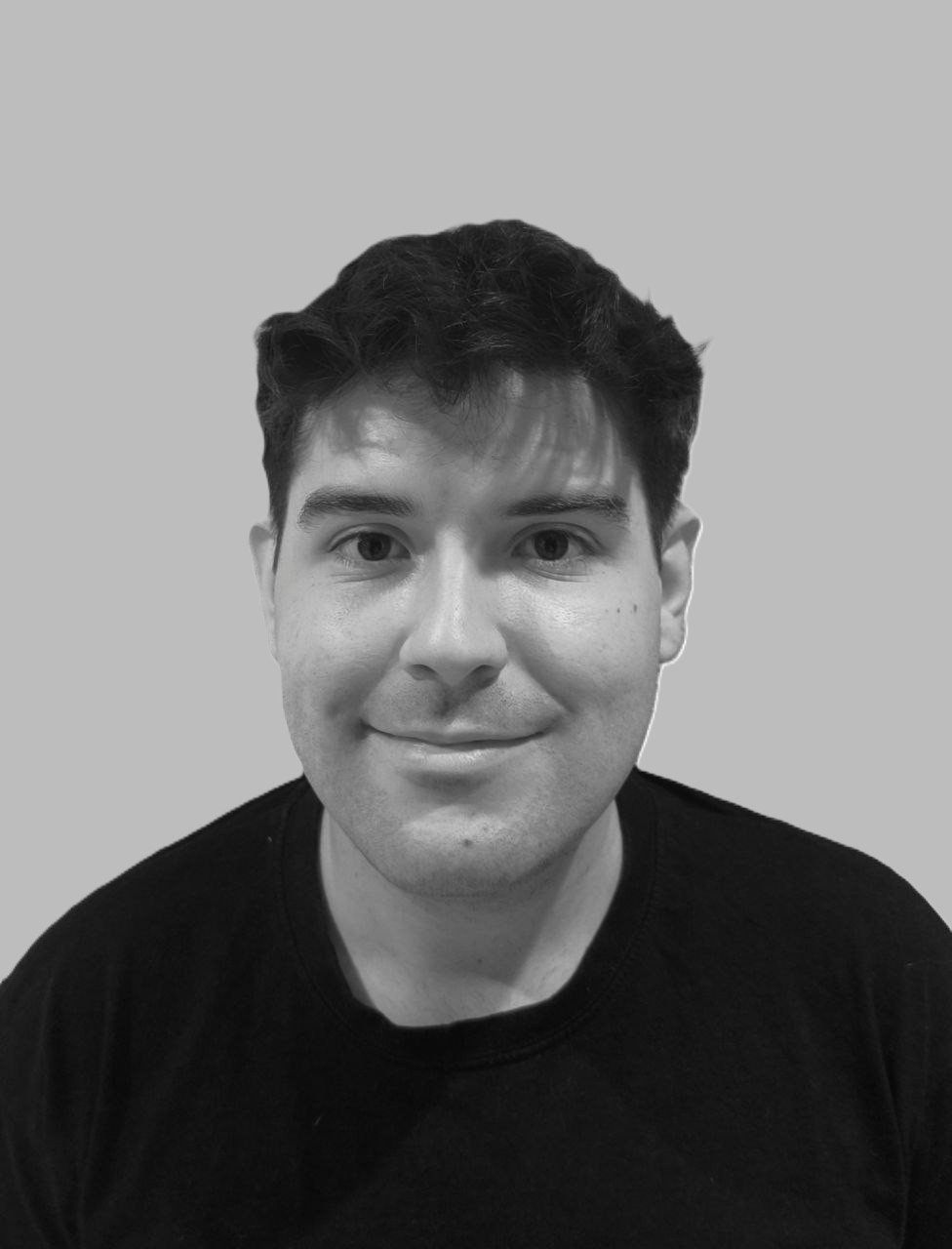}}] {Javier Fernandez} received the B.Sc. and the Ms degrees in Electronic engineering from Carlos III University, Madrid, Spain, in 2024 and 2025 respectively, where he is currently pursuing his Ph.D. His current interests are mixed-signal integrated circuit design and VCO-ADC design. 
\end{IEEEbiography}

\begin{IEEEbiography}[{\includegraphics[width=1in,height=1.25in,clip,keepaspectratio]{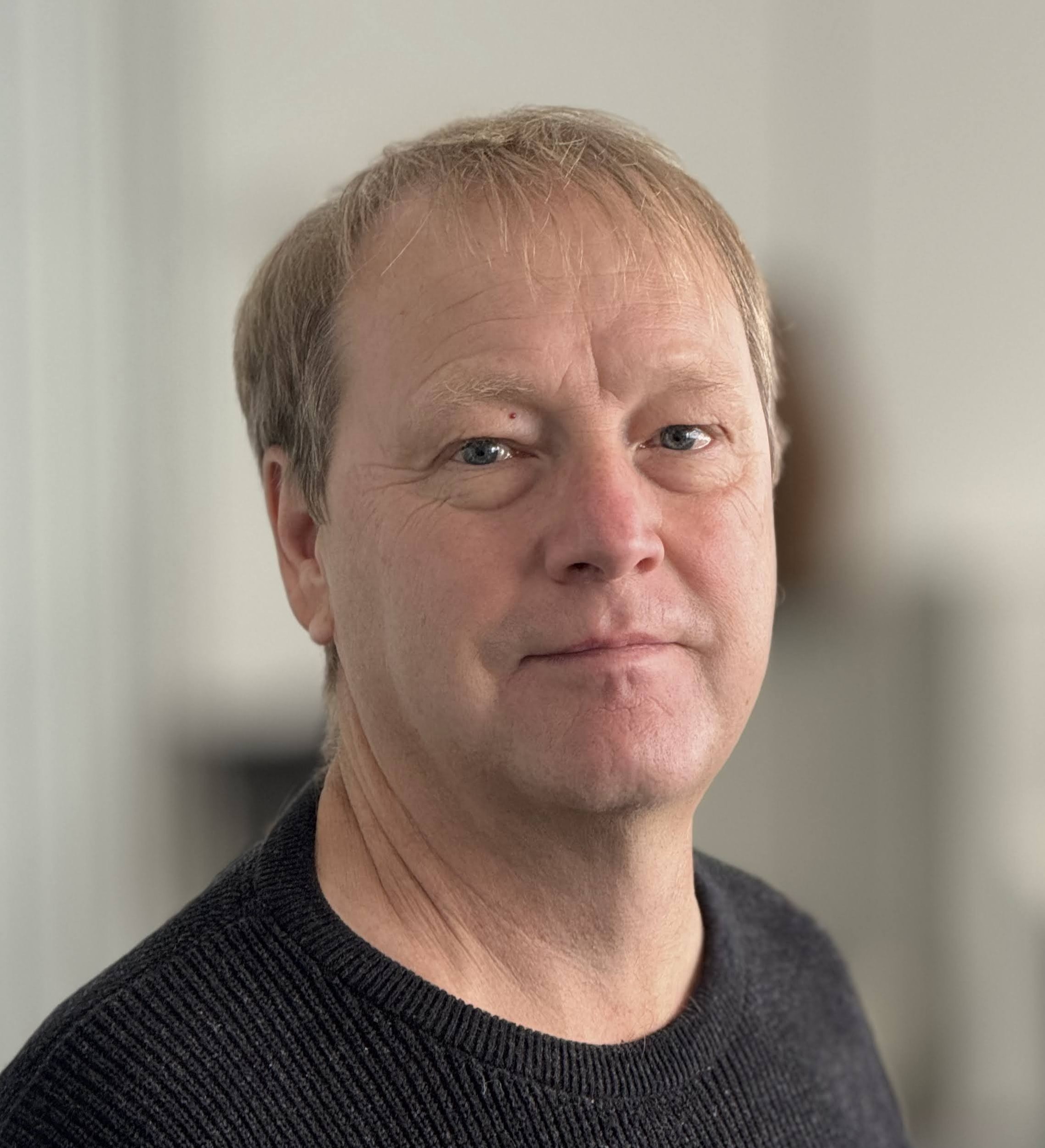}}] {Dietmar Straeussnigg} Dietmar Straeussnigg received the M.Sc. and Ph.D. degrees in electrical engineering and the Ph.D. degree in electrical engineering, from the University of Technology, Graz, Austria, in 1991 and 1995, respectively. In 1996, he joined the Siemens Microelectronics Design Center in Villach, Austria, which in 2000 became Infineon Technologies. From 1996 to 2005 he was concept engineer in the field of modem development (ADSL, VDSL). From 2005 to 2009 he was responsible for concept development for mixed-signal modules and digital signal processing. In 2009 he joined Lantiq Austria. In 2011 he went back to Infineon Technologies in the concept development group for silicon microphone products. In his current position, he is a Lead Principal Engineer, he is responsible for the development of advanced digital concepts for audio applications. Since 1998 he teaches as a part time lecturer for Digital Signal Processing, Digital Communications and Control Systems at the Carinthia University of Applied Sciences. He has contributed to more than 100 patents and has authored or coauthored to several journal and conference papers.
\end{IEEEbiography}

\begin{IEEEbiography}[{\includegraphics[width=1in,height=1.25in,clip,keepaspectratio]{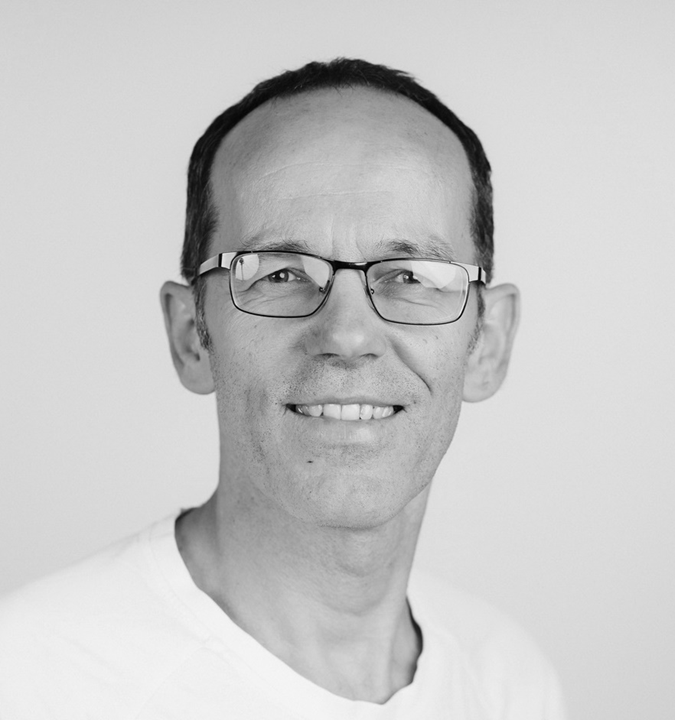}}] {Andreas Wiesbauer} (Member, IEEE) received the M.Sc. and Ph.D. degrees in electrical engineering from Vienna University of Technology, Vienna, Austria, in 1991 and 1994, respectively.,From 1996 to 1997, he was a Post-Doctoral Research Associate with Oregon State University, Corvallis, OR, USA, focusing on research of sigma–delta converters. In 1997, he joined the Siemens Microelectronics Design Center, Villach, Austria, which in 2000 became Infineon Technologies’ Design Center Austria. From 2000 to 2007, he was the Head of a design group of analog-/mixed-signal development for CMOS transceiver circuits. From 2007 to 2010, he was the Head of the RF-Power Design Team and did research in advanced RF power amplifier topologies. In his current position, he is a Distinguished Engineer of mixed-signal design and analog circuit concepts, with a focus on advanced audio applications, such as silicon microphones. During his academic and professional career, he has authored or coauthored several book chapters and more than 50 publications in international peer-reviewed journals and at international conferences. He has been granted over 80 patents, thereof more than 40 U.S. patents, in the field of analog-/mixed-signal design and data converter systems.,Dr. Wiesbauer served several years as a member for the ESSCIRC Technical Program Committee. 
\end{IEEEbiography}

\begin{IEEEbiography}[{\includegraphics[width=1in,height=1.25in,clip,keepaspectratio]{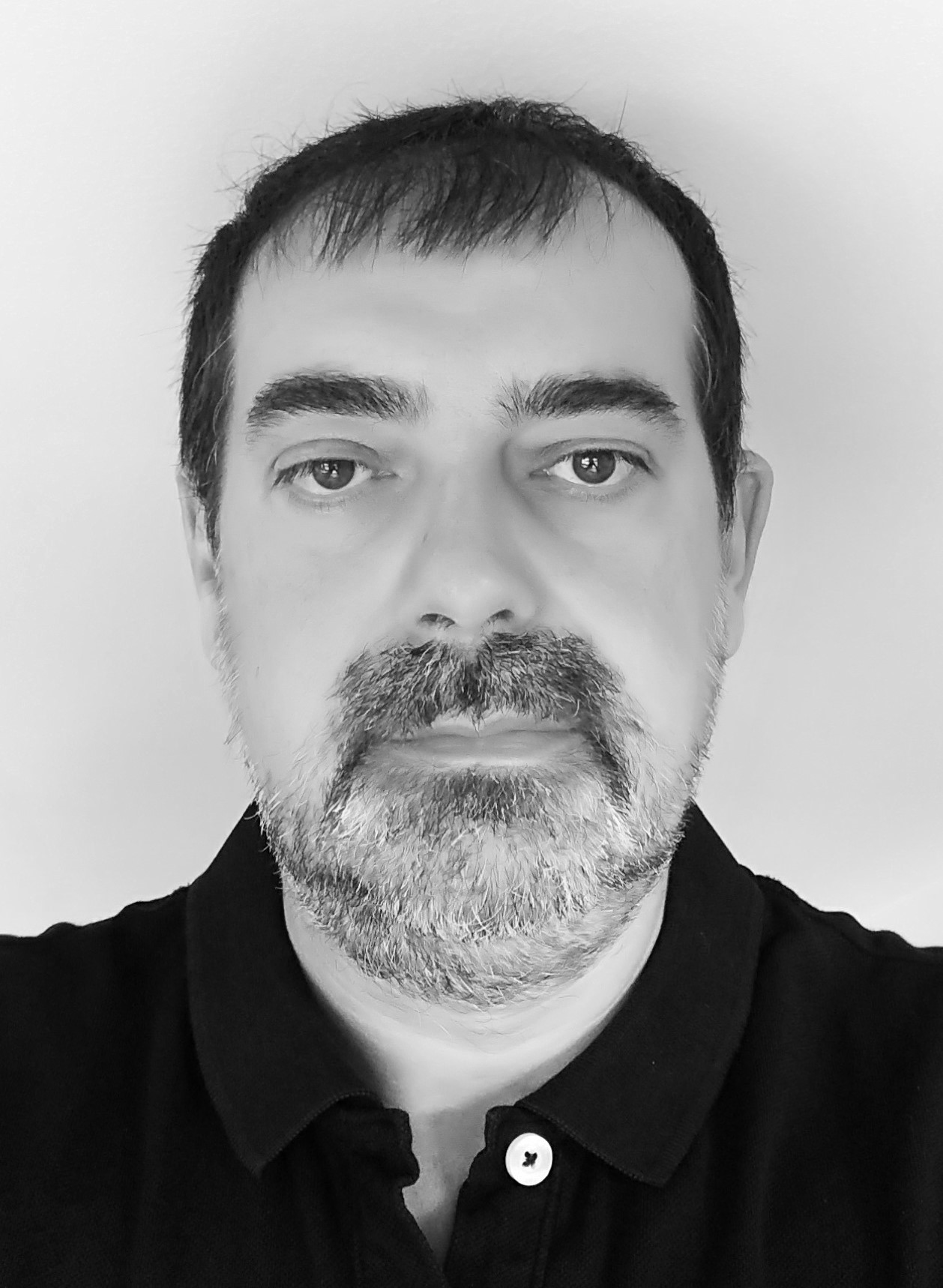}}] {Luis hernandez} (Senior Member, IEEE)  received the M.S. and Ph.D. degrees in telecommunication engineering from the Polytechnic University of Madrid in 1989 and 1995, respectively. He did a post-doctoral stay with the ECE Department, Oregon State University, in 1996, and Analog Devices, Willmington, USA, in 1997. In 1998, he joined the University Carlos III of Madrid, where he is currently a Full Professor with the Electronic Technology Department and leads the Mixed Signal Research Group. He has been the Department Head and the Ph.D. Program Director. In 2009, he did a sabbatical stay with IMEC, Leuven, Belgium. He has co-authored three books, over 200 papers and holds 27 patents.  He is a member of the IEEE-CAS ASPTC Committee. He has been an Associate Editor of IEEE Transactions on Circuits and Systems—I: Regular Papers and IEEE Transactions on Circuits and Systems—II: Express Briefs, for nine years and server for several years in the ESSCIRC program comitee. He is currently the deputy director of the Microelectronic Design Master program of the Carlos III University. His topics of interest are analog microelectronics, sigma-delta modulation, time-encoded data converters, and neural networks.
\end{IEEEbiography}

\end{document}